\documentclass[aip,cha,reprint,a4paper,nofootinbib]{revtex4-1}
\setcitestyle{numbers,square}

\usepackage[utf8]{inputenc}

\usepackage[pdftex]{graphicx}
\usepackage{subfigure}
\usepackage{amsmath, amssymb}
\usepackage{enumitem}
\usepackage{color, colortbl}
\usepackage{afterpage}
\usepackage{url}
\usepackage{multirow}
\usepackage{rotating}
\usepackage{scalefnt}
\usepackage{hyperref}
\usepackage{braket}
\usepackage{longtable}

\usepackage{soul}

\usepackage{braket,cleveref}

\usepackage{xr}

\begin{document}

\title{Network-based ranking in social systems: three challenges}

\author{Manuel S. Mariani}
\email{manuel.mariani@business.uzh.ch}
\affiliation{Institute of Fundamental and Frontier Sciences, University of Electronic Science and Technology of China, Chengdu 610054, PR China}
\affiliation{URPP Social Networks, University of Zurich, CH-8050 Zurich, Switzerland}

\author{Linyuan L{\"u}} 
\email{linyuan.lv@uestc.edu.cn}
\affiliation{Institute of Fundamental and Frontier Sciences, University of Electronic Science and Technology of China, Chengdu 610054, PR China}
\affiliation{Alibaba Research Center for Complexity Sciences, Hangzhou Normal University, Hangzhou 311121, PR China}
\affiliation{Beijing Computational Science
Research Center, Beijing 100193, PR China}

\begin{abstract} 
Ranking algorithms are pervasive in our increasingly digitized societies, with important real-world applications including recommender systems, search engines, and influencer marketing practices. 
From a network science perspective, network-based ranking algorithms solve fundamental problems related to the identification of vital nodes for the stability and dynamics of a complex system.
Despite the ubiquitous and successful applications of these algorithms, we argue that our
understanding of their performance and their applications to real-world problems face three fundamental challenges: (i) Rankings might be biased by various factors; (2) their effectiveness might be limited to specific problems; and (3) agents' decisions driven by rankings might result in potentially vicious feedback mechanisms and unhealthy systemic consequences.
Methods rooted in network science and agent-based modeling can help us to understand and overcome these challenges. 
\end{abstract}

\maketitle

\onecolumngrid

\section{Introduction}

The roots of research on ranking in social systems can be traced back to the 40s and 50s, when early studies introduced methods to quantify the social status of the members of a social system~\cite{seeley1949net,katz1953new,vigna2016spectral}.
Today, with the increasing availability of massive datasets on human activity, research on ranking algorithms for social systems is highly interdisciplinary, and it has significant real-world applications.
These two features are sharply exemplified by Google's PageRank. Introduced in 1998 by Brin and Page, this ranking algorithm became massively popular because of its implementation in Google's Web search engine~\cite{brin1998anatomy}, which has triggered a large wave of interest by computer scientists on its mathematical properties~\cite{berkhin2005survey} and applications to information retrieval problems~\cite{langville2011google}. At the same time, its defining equation had been already formulated in the social science literature in 1991~\cite{friedkin1991theoretical,friedkin2014two}, and the algorithm and its variants have found countless scientific applications beyond their original domain\cite{gleich2015pagerank},
including the quantification of scientific impact~\cite{zhou2012quantifying}, social leadership~\cite{lu2011leaders}, species importance for an ecosystem's stability~\cite{allesina2009googling}, and sport performance~\cite{radicchi2011best}. Real-world applications of ranking go far beyond Web search engines: every day, reputation systems estimate the trustworthiness of users and retailers in online marketplaces such as AirBnb~\cite{abrahao2017reputation}; automated recommender systems can be used to suggest new products to users in e-commerce platforms like Alibaba and Amazon~\cite{smith2017two}, or suitable investors for new startups~\cite{xu2020recommending}; rankings of prospect influencers can be used by online content creators and companies to detect the best candidates for product endorsements~\cite{lanz2019climb}.

These examples highlight the relevance of ranking various kinds of agents that compose social systems, including pieces of information (like websites and scientific papers), individuals, and businesses. A data-driven ranking of a given class of agents can be obtained through a variety of techniques, ranging from network-based algorithms (the main focus of this Perspective) to machine-learning algorithms (e.g., matrix factorization for recommender systems~\cite{koren2009collaborative} and latent semantic models to quantify the relevance of a document to a query~\cite{huang2013learning}) and methods based on dynamical models (e.g., fitness-based models~\cite{kong2008experience}).
Among all possible ranking techniques, network-based ranking algorithms play a prominent role~\cite{lu2016vital,liao2017ranking}. This class of algorithms take as input a given network representation of a complex system, and assign a score to each node meant to quantify its structural importance (or centrality) in the network. These algorithms can solve important problems related to the structure and dynamics of complex systems. For example, building on optimal percolation theory, the collective influence algorithm finds the minimal set of nodes that, when removed, cause the collapse of the giant component of the network, which has useful implications for information spreading, marketing campaigns, and immunization interventions~\cite{morone2015influence}. Under some assumptions about the topology of the network and the spreading dynamics, the nonbacktracking centrality quantifies the average size of the spreading processes initiated by a node~\cite{radicchi2016leveraging}, which might be informative for campaigns aimed to maximize the reach of information spreading. 


In this Perspective, we argue that although the solution to these problems indicates the importance and usefulness of network-based ranking algorithms, there exist three fundamental challenges faced by researchers interested in the development and application of network-based ranking algorithms for social systems: (1) \textit{Bias.} Rankings can be biased by various factors. For example, rankings of scientific papers based on citation networks can be biased by the papers' publication date, scientific field, and type of article.
(2) \textit{Performance variability.} A ranking algorithm's good performance in a given problem does not guarantee good performance in another problem (\textit{cross-problem performance variability}). Even when considering a particular problem, how the problem is specified can radically impact on the performance (\textit{within-problem performance variability}).
(3) \textit{Systemic consequences.} In social systems, ranking algorithms influence the behavior of the agents, which results in potentially vicious feedback mechanisms and unhealthy systemic consequences. 

The main goal of this Perspective is to provide explicit examples of these three threats to the design and application of ranking algorithms, and to suggest possible ways to overcome them. We anticipate one of the main takeaways of our Perspective: To properly counteract potential biases, performance limits and unhealthy systemic consequences of ranking algorithms, it is vital to understand the mechanisms behind the emergence of these three challenges.
This can be achieved by leveraging models that describe how a social system evolves over time, including models of growing social and information networks~\cite{medo2011temporal} and agent-based models where the components of the system behave and interact according to a set of rules~\cite{miller2009complex}.

\section{Network-based ranking algorithms}

In this Section, we provide a brief overview of network-based ranking algorithms~\cite{lu2016vital,liao2017ranking}.
For a simple network representation where nodes are connected by links, counting each node's number of connections provides us with the degree centrality -- arguably, the simplest proxy for the node's centrality~\cite{liao2017ranking}.
Starting from the degree, one can gradually incorporate higher-order network information by iteratively applying an operator to the vector of nodes' scores, which leads to the $H$-index and $k$-core centrality (or coreness)~\cite{lu2016h}. The $H$-index is an example of local centrality, meaning that the score of each node can be computed by only including information of the neighborhood of the node. Other prominent examples of local centralities include the aforementioned collective influence metric~\cite{morone2015influence}, and the LocalRank centrality, a heuristic metric to identify influential spreaders of information~\cite{chen2012identifying}.

The eigenvector, Katz, PageRank, LeaderRank, nonbacktracking centralities are all examples of global ranking algorithms based on the solution of an eigenvalue problem, and most of these ``spectral ranking" techniques can be also interpreted as weighted combinations of network paths~\cite{vigna2016spectral,liao2017ranking}. Another important class of algorithms builds on pairwise network distances. The average distance of a node from the other nodes in the network can be interpreted as its centrality. Different choices of the distance function lead to different centrality metrics: for example, the shortest-path distance leads to the classical closeness centrality, whereas the random-walk effective distance leads to the ViralRank centrality~\cite{iannelli2018influencers}.
Another classical algorithm is the betweenness centrality and its variants, which build on the assumption that a given node $i$ is central if many shortest paths that connect pairs of nodes pass through node $i$~\cite{brandes2001faster}.

Finally, an important class of network-based ranking algorithms aims to order the nodes of a system based on the results of (potentially incomplete or noisy) observed sets of pairwise interactions~\cite{bradley1952rank,fogel2016spectral,cucuringu2016sync,de2018physical,d2019ranking}. These algorithms seek to find a permutation of the nodes that optimizes a predefined cost function~\cite{de2018physical}. From this perspective, the edges of a network are interpreted as the consequence of hidden hierarchies and, therefore, one can attempt to infer the underlying hierarchies from the observed outcomes.
This family of algorithms typically applies to directed weighted networks, with a broad spectrum of potential applications, ranging from inferring a ranking of sport players/teams in tournaments from the results of their matches~\cite{fogel2016spectral}, to inferring a ranking of universities from the mobility of Ph.~D. graduates~\cite{de2018physical}. 
Different cost functions have been studied in the literature, leading to various algorithms, e.g., SerialRank~\cite{fogel2016spectral}, SyncRank~\cite{cucuringu2016sync}, and SpringRank~\cite{de2018physical}.

We refer the interested reader to~\cite{lu2016vital,liao2017ranking} for detailed reviews on ranking in complex networks, including formulations of centrality for more complex network representations (like weighted networks, higher-order networks, multilayer networks).
Crucially, each centrality metric builds on a specific assumption on what being an important node means. While the 
algorithms' assumptions and their translations into mathematical equations are generally plausible, important challenges arise when applying the resulting ranking algorithm to specific problems, which is the focus of the following sections.

\section{Bias}

 \begin{figure*}[t]
        \centering
        \includegraphics[scale=0.35]{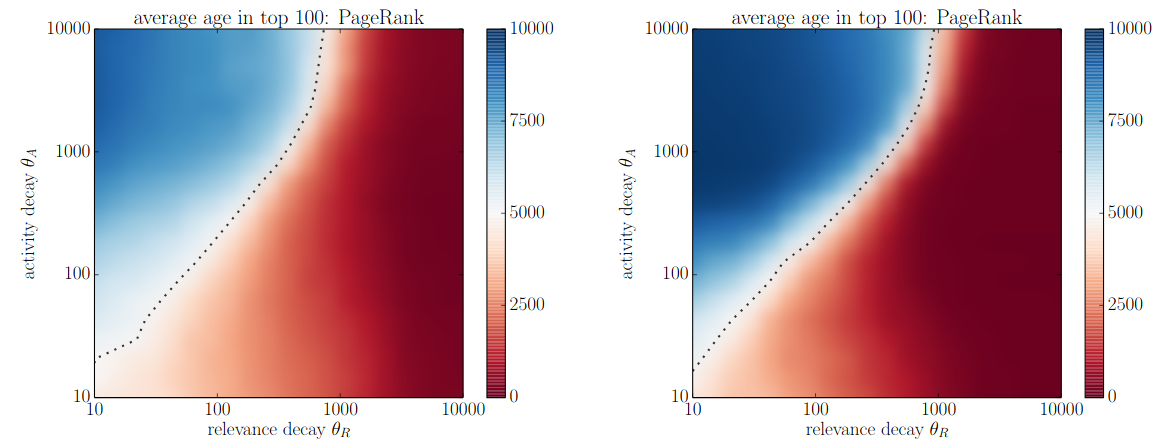}
        \caption{\textbf{Bias: the case of PageRank's age bias.} We focus on growing directed networks of $10,000$ nodes where each node can receive incoming edges or create outgoing edges. Nodes are labeled by entrance order ($i\in\{1,\dots,10,000\}$, where node $1$ and $10,000$ are the first and last nodes to enter the network, respectively). The two panels show the average entrance order of the top $100$ nodes in the ranking by PageRank, $\tau_{100}$, in networks generated with two different models. Both models are characterized by two main control parameters: the timescale of relevance decay, $\Theta_R$ (i.e., the decay of node likelihood to attract incoming edges), and activity decay, $\Theta_A$ (i.e., the decay of node likelihood to create outgoing edges). The ranking by PageRank is (nearly) unbiased only over a narrow region of the parameter space where $
        \tau_{100}\simeq 5,000=N/2$. When the relevance decay is slower than the activity decay ($\Theta_R\gg\Theta_A$), PageRank tends to be biased toward old nodes ($\tau_{100}<N/2$). A bias toward recent nodes ($
        \tau_{100}>N/2$) emerges instead when the relevance decay is faster than the activity decay ($\Theta_R\ll\Theta_A$).
        Reprinted from~\cite{mariani2015ranking}.}
        \label{fig:bias}
    \end{figure*}

The main hope motivating the use of algorithms for ranking and prediction tasks is that they might provide an objective estimation of the value of an agent (whether the quality of a cultural product, the talent of an individual, or the relevance of a webpage), whereas human or expert judgment might be subjective and influenced by biases and social factors. 
Besides, algorithms can crunch massive datasets in relatively little time, allowing for fast retrieval of relevant information in a short time. However, accepting blindly the outcomes by a ranking algorithm is potentially misleading, and we argue here that an appropriate dose of caution is necessary when interpreting the results by a given algorithm as a signal of quality or talent. An important shortcoming of ranking algorithms is that -- in a similar way as machine learning algorithms~\cite{ntoutsi2020bias} -- they can be biased by multiple confounding factors.

Consider the popular Google PageRank algorithm, which builds on a diffusion process on a directed network where each node can have incoming or outgoing edges. In a citation network, for example, a scientific paper's outgoing and incoming edges represent its references to other papers and its received citations from other papers, respectively. PageRank assigns to each node a score equal to the stationary probability of a stochastic process that combines a random walk along the network's edges with a random teleportation mechanism~\cite{gleich2015pagerank}. 
The algorithm reflects the plausible assumption that a node is important if other important nodes endorse it -- e.g., in a citation network, a paper is important if other important papers cite it.
However, consider now the first direct experimental observation of gravitational waves and the Physical Review Letter paper that presented this discovery~\cite{abbott2016observation}, published in February 2016. Were we to blindly trust Google’s PageRank, at the end of 2016, we would conclude that among all the papers in the APS corpus, this letter deserves the $12,482$nd place by importance. Instead, even though it is impossible to establish the exact ranking position deserved by the paper, virtually nobody would doubt that the the paper's finding constitutes one of the main milestones in the history of physics.
The crux of the problem lies in a fundamental bias: in first approximation, the PageRank score of a paper is expected to be linearly correlated with its number of received citations~\cite{fortunato2006approximating} and, therefore, it strongly favors old nodes over recent ones~\cite{mariani2015ranking}. 

Stochastic models of network growth help us to understand why the biases of ranking algorithms emerge~\cite{mariani2015ranking}. One can indeed grow synthetic directed networks where the probability that a node receives new incoming edges is proportional to its number of previous links (preferential attachment), to its intrinsic fitness, and a time-decaying function that represents the node's relevance decay over time~\cite{medo2011temporal}. Besides, existing nodes can activate to create outgoing edges at a rate that decays with their age~\cite{mariani2015ranking}.
This framework reveals that the interplay between the timescales of the two time-decay processes leads to the PageRank's temporal bias (Fig.~\ref{fig:bias}): when the aging of relevance is slower (faster) than the decay of activity, most directed links point from a node to an older (more recent) node, causing the diffusion process that determines the PageRank scores to favor old (recent) nodes~\cite{mariani2015ranking}.

While based on a simple model, this approach reveals that in growing networks, a large part of the variation in PageRank score is due to temporal effects. This suggests that PageRank's temporal bias can be removed by rescaling the scores with a transformation that ensures that the average score of the nodes and its standard deviation are independent of node age~\cite{mariani2016identification}. When such a transformation is applied, the resulting ``rescaled" score can detect much earlier important nodes, with useful implications for the early detection of milestone papers~\cite{mariani2016identification,xu2020unbiased}, patents~\cite{mariani2019early,xu2020unbiased}, and movies~\cite{ren2018randomizing}. The benefit from this procedure is exemplified, again, by the paper that reported the first direct observation of gravitational waves~\cite{abbott2016observation}: the paper is ranked $16$th by rescaled PageRank at the end of 2016, which constitutes a substantial improvement compared to the $12,482$nd position by the original PageRank, and suggests that the paper deserves a place among the most significant ones in the APS corpus.

While we have focused on the bias by node age of ranking algorithms, biases can emerge along multiple dimensions. For example, citation-based rankings of papers' scientific impact are biased by paper age and scientific field~\cite{waltman2016review,battiston2019taking}, which motivates the question whether both biases can be removed by a simple rescaling procedure~\cite{vaccario2017quantifying}. 
Existing results indicate that while simple rescaling procedures can dramatically mitigate the biases of a metric, it remains challenging to achieve a ranking that is statistically consistent with a ranking that is unbiased by construction, obtained by uniformly sampling the top-papers from different age and field groups~\cite{vaccario2017quantifying}. Similar age and field biases, and attempts to mitigate them, exist in rankings of researchers~\cite{dunaiski2019globalised}.

Subtler forms of bias can exist. In science, the number of citations received by a paper might not simply result from the intrinsic importance of the paper, but heavily depend on social mechanisms.
Social influence and reputation factors that may affect a paper's impact metrics include the authors' previous number of citations~\cite{petersen2014reputation} and centrality in the co-authorship network~\cite{sarigol2014predicting}, and reciprocity effects~\cite{li2019reciprocity}. Therefore, social mechanisms might affect rankings by impact and popularity, potentially mismatching rankings and the intrinsic value of the papers. Given the importance of impact indicators for academic careers, additional research is needed to understand whether it is necessary to factor out social effects from these metrics, and if yes, how to do so.

\section{Performance variability}

 \begin{figure*}[t]
        \centering
        \includegraphics[scale=0.275]{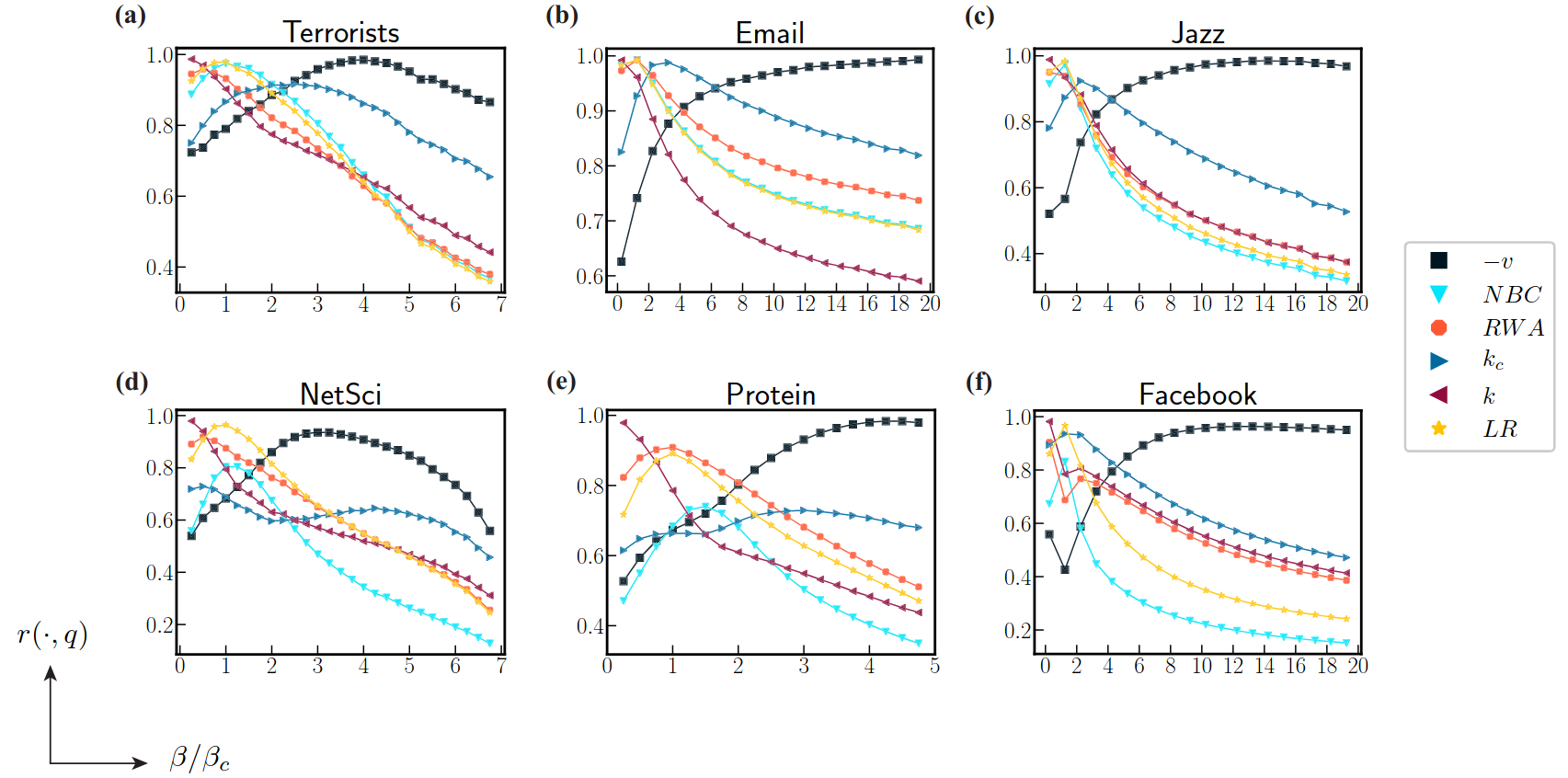}
        \caption{\textbf{Within-problem performance variability: the case of influential spreaders.} The performance of six different ranking algorithms (ViralRank, nonbacktracking centrality, random-walk accessibility, k-core centrality, degree, and LocalRank, see~\cite{iannelli2018influencers} for the definitions) is plotted against the SIR spreading model's transmission probability, $\beta$, for six empirical datasets -- see~\cite{iannelli2018influencers} for details. The performance of a metric is defined as the linear correlation between the node score they produce and the nodes' spreading ability. The spreading ability of a given node is defined as the average size of the spreading processes initiated by that node, and it is measured from the results of numerical simulations with the SIR model with a predefined transmission rate, $\beta$. The result shows that the algorithms' performance is highly dependent on $\beta$, meaning that a metric that performs well around the critical point of the model ($\beta/\beta_c\simeq 1$) can perform poorly well above the critical point ($\beta\gg \beta_c$). Reprinted from~\cite{iannelli2018influencers}.}
        \label{fig:performance}
    \end{figure*}

How to evaluate the performance of a ranking algorithm? In network science, ranking algorithms are routinely evaluated according to their ability to single out vital nodes for the system's structure and dynamics. From a structural perspective, rankings can enable the identification of a set of nodes that maximize a function of the structure of the network~\cite{lu2016vital}, e.g., the minimal set of \textit{structural influencers} whose removal causes the collapse of the network's giant component~\cite{morone2015influence}.
 From a dynamical perspective, rankings can enable the identification of a set of nodes that maximize a function of the structure and dynamics of the network (\textit{functional influence maximization}~\cite{lu2016vital}) -- e.g., a set of \textit{influential spreaders} that, when activated, trigger the largest cascade processes under a predefined spreading model~\cite{kitsak2010identification}. The importance of a set of nodes for dynamical processes on networks can be also assessed for synchronization processes, by quantifying the ability of a set of nodes to drive the system from an initial to a desired state in a short time~\cite{fan2020characterizing}. 
 
One can also evaluate ranking algorithms based on their performance in various kinds of predictive problems. In the machine-learning literature, an important stream of works~\cite{fogel2016spectral, cucuringu2016sync,de2018physical,d2019ranking} have assessed the rankings' ability to predict the outcomes of pairwise interactions. Such predictive power can be tested in empirical data on sport tournament results, animal dominance, and faculty hiring, among others~\cite{de2018physical}.
 In science of science, rankings can be evaluated according to their ability to early detect small sets of scientific papers or patents that have been labeled by experts as groundbreaking or seminal~\cite{mariani2016identification,mariani2019early}.

A key challenge is that the performance of an algorithm in one of these problems does not predict its performance in another one (\textit{cross-problem variability}).
For example, by comparing 17 network-based ranking algorithms, a recent study~\cite{xu2020unbiased} found that time-rescaled versions of PageRank and its variant LeaderRank~\cite{lu2011leaders} are the best-performing algorithms in the identification of expert-selected seminal papers and patents. PageRank is also effective in identifying influential researchers~\cite{dunaiski2018author}. However, PageRank performs poorly in other problems. Other centrality metrics -- like the degree centrality, the k-core centrality~\cite{kitsak2010identification}, and ViralRank~\cite{iannelli2018influencers} -- substantially outperform PageRank in finding the most influential spreaders in a network~\cite{iannelli2018influencers}. Building on optimal percolation theory, the collective influence metric significantly outperforms PageRank in detecting the structural influencers~\cite{morone2015influence}. 

But even when considering a single problem, how the problem is specified can substantially alter the results (\textit{within-problem variability}). For example, in the problem of identifying the influential spreaders based on the SIR model, the algorithms' performance depends widely on the transmission probability, a key parameter of the spreading model~\cite{iannelli2018influencers} (Fig.~\ref{fig:performance}), on the type of dynamics in exam (e.g., contact vs. reaction-diffusion dynamics)~\cite{iannelli2018influencers}, and on the timescale at which the dynamics is observed~\cite{zhou2019fast}.
The main takeaway from these studies is that there might exist no all-weather algorithms. The good performance of an algorithm in a given problem does not guarantee good performance in another problem, and even for the same problem, different problem specifications might require different algorithms. This uncertainty highlights the importance of reporting clearly the scenarios where an algorithm is effective, and the scenarios where the algorithm fails. In the problem of identifying the influential spreaders under a given spreading model, this can be achieved by reporting the algorithms' performance over a broad range of values of the model parameters, and understanding which parameter ranges are the most relevant ones for various real-world processes.

The choice of the performance evaluation metric is critical as well. In the problem of identifying expert-selected important nodes (papers or patents) in science and technology, the age distribution of the expert-selected nodes can significantly impact on the algorithms' performance: if the expert-selected nodes are old ones, performance evaluation metrics that ignore this bias will favor ranking algorithms that are biased in favor of old nodes~\cite{mariani2016identification,dunaiski2019interplay,xu2020unbiased}. ``Corrected" performance evaluation metrics that penalize biased metrics are not affected by this confounding effect~\cite{xu2020unbiased}. However, there is not yet a unique and universally-agreed way to evaluate ranking algorithms for scientific and technological impact. Given the role played by impact metrics in research evaluations~\cite{rijcke2016evaluation}, we need a deeper understanding of their biases and their ability to quantify productivity, talent, and impact. Toward this direction, the introduction of a ``gold standard" for the evaluation of impact metrics for academic actors at various levels -- from researchers to departments and universities -- is highly desirable.

\section{Systemic consequences}

 \begin{figure*}[t]
        \centering
        \includegraphics[scale=0.45]{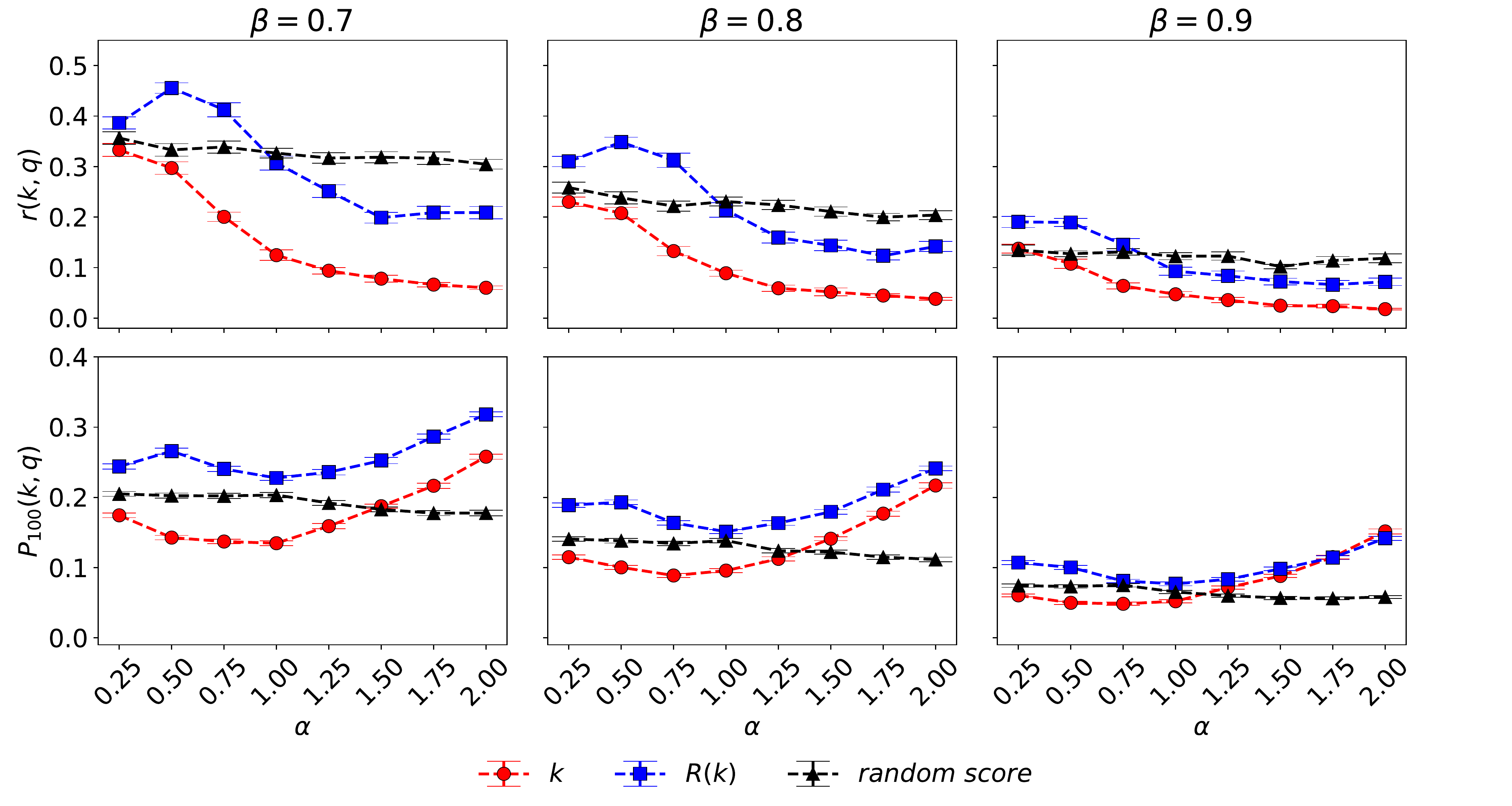}
        \caption{\textbf{Systemic consequences: the correlation between popularity and quality.} Results on growing network models where nodes are driven both by ranking and by the intrinsic fitness of the other nodes~\cite{zhang2019long}. Control parameter $\beta$ determines the relative importance of ranking and fitness, whereas parameter $\alpha$ determines how sensitive the agents are to the other nodes' ranking position. When the nodes follow the ranking by age-rescaled popularity, $R(k)$, the correlation between node popularity and intrinsic fitness, $r(k,q)$, is significantly larger than when the nodes follow the ranking by popularity, $k$ (top panels). A qualitatively similar result holds for the overlap between the top-$100$ nodes by popularity and quality, $P_{100}(k,q)$ (bottom panels). Surprisingly, when agents are highly sensitive to ranking ($\alpha>1$, roughly), all popularity-based metrics lead to network with a lower popularity-quality correlation than networks where agents follow a random ranking (black lines). Adapted from~\cite{zhang2019long}.}
        \label{fig:consequences}
    \end{figure*}

A focus on bias removal and performance alone might miss an important property which makes social systems fundamentally different from physical systems composed of atoms and molecules.
Rankings can indeed alter how the members of a social system behave: As ranking algorithms are adopted and used in a social system, they can influence the behavior of the agents, which in turn influences the rankings themselves. For example, experimental studies on cultural markets found that when the members of a cultural market are aware of the rankings by popularity of cultural products, the system exhibits wider popularity inequalities compared to conditions where the members are unaware of rankings~\cite{salganik2006experimental}. 
Can we predict how the adoption of an algorithm in a given system will alter the agents' behavior and further influence the evolution of the system?
Most studies that aimed to answer this question have focused on agent-based models and network formation models~\cite{bardoscia2013social,zeng2015modeling,zhang2019long,livan2019don,d2020fairness}.
 In the following, we highlight three examples of insights that can be gained from stochastic models of network formation.

First, in a social network, when agents strive to connect to high-ranked agents (i.e., central agents) and delete their links to low-ranked agents (i.e., peripheral agents), highly-hierarchical societies emerge, with reduced social mobility~\cite{konig2011network,bardoscia2013social,konig2014nestedness,mariani2019nestedness}. From a statistical physics perspective, one can formulate a growing network model where, at each step, a selected agent (with probability $\alpha$) creates a connection to the most central agent they are not connected to, or (with probability $1-\alpha$) deletes its connection to the least-central agent among its neighbors. The model features a phase transition at a critical value $\alpha=\alpha_c$ which separates a phase where the resulting network is fully-connected ($\alpha>\alpha_c$) from a phase where the network exhibits a highly centralized topology ($\alpha<\alpha_c$), namely a perfectly nested one where agents' interaction are hierarchically arranged~\cite{mariani2019nestedness}. Importantly, this result holds regardless of the algorithm employed by the agents to assess the centrality of the other agents~\cite{konig2011network}. Results on systems composed of a constant number of agents indicate that highly-hierarchical structures are associated with reduced social mobility, meaning that it is harder for agents to improve their ranking in society~\cite{bardoscia2013social}.

Second, in a growing network, if the new nodes choose their connections driven by a ranking metric that is biased by node age, the resulting system will exhibit uneven popularity distribution, and the overall correlation between talent and success will be low~\cite{zhang2019long}. This property can be observed in a growing network model where, when establishing new links, a node is (with probability $\beta$) driven by the ranking position of the preexisting nodes, or (with probability $1-\beta$) by their intrinsic fitness -- a parameter that quantifies the inherent appeal of the node. When the node chooses by ranking, the probability that it chooses node $j$ decays with the ranking position $r_j$ of the node as $r_j^{-\alpha}$, where $
\alpha$ and $\beta$ are control parameters of the model. The model can be used to grow synthetic networks under the influence of different ranking algorithms and thus, to compare the long-term implications of the ranking by a metric that is biased by node age (e.g., the total number of links received by the node) against those by a metric that is not biased (e.g., a time-rescaled link count). Numerical simulations indicate that when the nodes follow the unbiased metric, the overall correlation between the final popularity of the nodes and their intrinsic fitness is significantly higher (Fig.~\ref{fig:consequences}), and popularity is more evenly distributed~\cite{zhang2019long}.

Third, if the agents attempt to mimic the strategies of the highest-ranked individuals, the overall welfare of the society increases, but so do success inequalities~\cite{livan2019don}.
This has been demonstrated with a model where a given agent benefits from a given action due to both the intrinsic payoff associated with the action and her ability to produce a successful outcome out of it. Each agent starts with a random action, and at each time step of the dynamics, each agent has the opportunity to change action. With a probability $q$, the agent copies the action of a randomly-selected agent among those ranked better than her (\textit{imitation}), whereas with a probability $1-q$, she selects a random action (\textit{serendipity}). The control parameter $q\in[0,1]$ determines the relative importance of imitation and serendipity for agents' choices. 
As $q$ increases, the total welfare of the system increases, but so do success inequalities, whereas the correlation between agents' success and talent decreases~\cite{livan2019don}.

Taken together, these studies indicate that agent-based models can shed light on the possible systemic consequences of ranking algorithms on a given system. In particular, agent-level actions motivated by the results of ranking algorithms can result in unhealthy systemic consequences, like reduced social mobility~\cite{bardoscia2013social} and low correlation between success and merit~\cite{zhang2019long,livan2019don}.
In future research, we expect investigations of more complex models to deepen our understanding of the impact of rankings on the structure and stability of social systems. Beyond extensive explorations of various theoretical scenarios, we also expect future research to put more emphasis on the calibration of the models on empirical data: once the agents' behavior is understood through laboratory experiments or observational data analysis, the systemic consequences of the observed micro-level behavior can be grasped through agent-based models calibrated on empirical behavior~\cite{smith2018simulating,schweitzer2018sociophysics}.

\section{Conclusion}

The promise that algorithms deliver objective estimations of quality, talent, and importance is challenged by the algorithms' potential biases, performance variability, and systemic consequences.
We argue that these challenges need to be carefully examined by researchers who aim to develop new ranking algorithms, researchers interested in applying an existing ranking algorithm to answer a given research question, and by policymakers who aim to apply quantitative methods to assess an individual's or organization's talent and their likelihood of future success.
Although we have focused on network-based ranking algorithms, these challenges potentially apply to any ranking method for a social system. This is exemplified by recent cross-disciplinary efforts  to understand, quantify, and mitigate the bias of machine-learning algorithms employed by governments and organizations~\cite{ntoutsi2020bias}, and to predict the long-term consequences of such biases through computer simulations~\cite{d2020fairness}.

We have focused on these three challenges because of their potential interest to the physics and complexity science community.
Of course, other challenges exist and need attention by the scientific community, including the algorithms' computational complexity, the dependence of their outcomes (and biases) on data quality and completeness, ethical concerns related to their real-world application, and their robustness against malicious attacks and manipulation, among others. 
We hope that this Perspective will increase the awareness of the potential limitations of ranking algorithms and inspire studies aimed to overcome them.

\section*{Acknowledgements}

We thank Shuqi Xu for providing us the rankings of papers in the APS corpus at the end of 2016, and Shilun Zhang for providing us Figure 3. We thank Flavio Iannelli, Matúš Medo, and Giacomo Vaccario for their detailed feedback which helped us to improve the text. This work was supported by the National Natural Science Foundation of China (Grants Nos. 61673150, 11622538), the Science Strength Promotion Program of the University of Electronic Science and Technology of China (Grant No. Y030190261010020), and the Natural Science Foundation of Zhejiang Province of China (Grant no. LR16A050001). MSM acknowledges financial support from the University of Zurich through the URPP Social Networks, the Swiss National Science Foundation (Grant No. 200021-182659), the UESTC professor research start-up (Grant No. ZYGX2018KYQD215).

\section*{Data availability statement}

Data sharing is not applicable to this article as no new data
were created or analysed in this study.

\section*{References}

\bibliography{bibliography}

\begin{thebibliography}{67}%
\makeatletter
\providecommand \@ifxundefined [1]{%
 \@ifx{#1\undefined}
}%
\providecommand \@ifnum [1]{%
 \ifnum #1\expandafter \@firstoftwo
 \else \expandafter \@secondoftwo
 \fi
}%
\providecommand \@ifx [1]{%
 \ifx #1\expandafter \@firstoftwo
 \else \expandafter \@secondoftwo
 \fi
}%
\providecommand \natexlab [1]{#1}%
\providecommand \enquote  [1]{``#1''}%
\providecommand \bibnamefont  [1]{#1}%
\providecommand \bibfnamefont [1]{#1}%
\providecommand \citenamefont [1]{#1}%
\providecommand \href@noop [0]{\@secondoftwo}%
\providecommand \href [0]{\begingroup \@sanitize@url \@href}%
\providecommand \@href[1]{\@@startlink{#1}\@@href}%
\providecommand \@@href[1]{\endgroup#1\@@endlink}%
\providecommand \@sanitize@url [0]{\catcode `\\12\catcode `\$12\catcode
  `\&12\catcode `\#12\catcode `\^12\catcode `\_12\catcode `\%12\relax}%
\providecommand \@@startlink[1]{}%
\providecommand \@@endlink[0]{}%
\providecommand \url  [0]{\begingroup\@sanitize@url \@url }%
\providecommand \@url [1]{\endgroup\@href {#1}{\urlprefix }}%
\providecommand \urlprefix  [0]{URL }%
\providecommand \Eprint [0]{\href }%
\providecommand \doibase [0]{http://dx.doi.org/}%
\providecommand \selectlanguage [0]{\@gobble}%
\providecommand \bibinfo  [0]{\@secondoftwo}%
\providecommand \bibfield  [0]{\@secondoftwo}%
\providecommand \translation [1]{[#1]}%
\providecommand \BibitemOpen [0]{}%
\providecommand \bibitemStop [0]{}%
\providecommand \bibitemNoStop [0]{.\EOS\space}%
\providecommand \EOS [0]{\spacefactor3000\relax}%
\providecommand \BibitemShut  [1]{\csname bibitem#1\endcsname}%
\let\auto@bib@innerbib\@empty
\bibitem [{\citenamefont {Seeley}(1949)}]{seeley1949net}%
  \BibitemOpen
  \bibfield  {author} {\bibinfo {author} {\bibfnamefont {J.~R.}\ \bibnamefont
  {Seeley}},\ }\bibfield  {title} {\enquote {\bibinfo {title} {The net of
  reciprocal influence. a problem in treating sociometric data},}\ }\href@noop
  {} {\bibfield  {journal} {\bibinfo  {journal} {Canadian Journal of
  Experimental Psychology}\ }\textbf {\bibinfo {volume} {3}},\ \bibinfo {pages}
  {234} (\bibinfo {year} {1949})}\BibitemShut {NoStop}%
\bibitem [{\citenamefont {Katz}(1953)}]{katz1953new}%
  \BibitemOpen
  \bibfield  {author} {\bibinfo {author} {\bibfnamefont {L.}~\bibnamefont
  {Katz}},\ }\bibfield  {title} {\enquote {\bibinfo {title} {A new status index
  derived from sociometric analysis},}\ }\href@noop {} {\bibfield  {journal}
  {\bibinfo  {journal} {Psychometrika}\ }\textbf {\bibinfo {volume} {18}},\
  \bibinfo {pages} {39--43} (\bibinfo {year} {1953})}\BibitemShut {NoStop}%
\bibitem [{\citenamefont {Vigna}(2016)}]{vigna2016spectral}%
  \BibitemOpen
  \bibfield  {author} {\bibinfo {author} {\bibfnamefont {S.}~\bibnamefont
  {Vigna}},\ }\bibfield  {title} {\enquote {\bibinfo {title} {Spectral
  ranking},}\ }\href@noop {} {\bibfield  {journal} {\bibinfo  {journal}
  {Network Science}\ }\textbf {\bibinfo {volume} {4}},\ \bibinfo {pages}
  {433--445} (\bibinfo {year} {2016})}\BibitemShut {NoStop}%
\bibitem [{\citenamefont {Brin}\ and\ \citenamefont
  {Page}(1998)}]{brin1998anatomy}%
  \BibitemOpen
  \bibfield  {author} {\bibinfo {author} {\bibfnamefont {S.}~\bibnamefont
  {Brin}}\ and\ \bibinfo {author} {\bibfnamefont {L.}~\bibnamefont {Page}},\
  }\bibfield  {title} {\enquote {\bibinfo {title} {The anatomy of a large-scale
  hypertextual {W}eb search engine},}\ }\href@noop {} {\bibfield  {journal}
  {\bibinfo  {journal} {Computer Networks and ISDN Systems}\ }\textbf {\bibinfo
  {volume} {30}},\ \bibinfo {pages} {107--117} (\bibinfo {year}
  {1998})}\BibitemShut {NoStop}%
\bibitem [{\citenamefont {Berkhin}(2005)}]{berkhin2005survey}%
  \BibitemOpen
  \bibfield  {author} {\bibinfo {author} {\bibfnamefont {P.}~\bibnamefont
  {Berkhin}},\ }\bibfield  {title} {\enquote {\bibinfo {title} {A survey on
  pagerank computing},}\ }\href@noop {} {\bibfield  {journal} {\bibinfo
  {journal} {Internet Mathematics}\ }\textbf {\bibinfo {volume} {2}},\ \bibinfo
  {pages} {73--120} (\bibinfo {year} {2005})}\BibitemShut {NoStop}%
\bibitem [{\citenamefont {Langville}\ and\ \citenamefont
  {Meyer}(2011)}]{langville2011google}%
  \BibitemOpen
  \bibfield  {author} {\bibinfo {author} {\bibfnamefont {A.~N.}\ \bibnamefont
  {Langville}}\ and\ \bibinfo {author} {\bibfnamefont {C.~D.}\ \bibnamefont
  {Meyer}},\ }\href@noop {} {\emph {\bibinfo {title} {Google's PageRank and
  beyond: The science of search engine rankings}}}\ (\bibinfo  {publisher}
  {Princeton university press},\ \bibinfo {year} {2011})\BibitemShut {NoStop}%
\bibitem [{\citenamefont {Friedkin}(1991)}]{friedkin1991theoretical}%
  \BibitemOpen
  \bibfield  {author} {\bibinfo {author} {\bibfnamefont {N.~E.}\ \bibnamefont
  {Friedkin}},\ }\bibfield  {title} {\enquote {\bibinfo {title} {Theoretical
  foundations for centrality measures},}\ }\href@noop {} {\bibfield  {journal}
  {\bibinfo  {journal} {American journal of Sociology}\ }\textbf {\bibinfo
  {volume} {96}},\ \bibinfo {pages} {1478--1504} (\bibinfo {year}
  {1991})}\BibitemShut {NoStop}%
\bibitem [{\citenamefont {Friedkin}\ and\ \citenamefont
  {Johnsen}(2014)}]{friedkin2014two}%
  \BibitemOpen
  \bibfield  {author} {\bibinfo {author} {\bibfnamefont {N.~E.}\ \bibnamefont
  {Friedkin}}\ and\ \bibinfo {author} {\bibfnamefont {E.~C.}\ \bibnamefont
  {Johnsen}},\ }\bibfield  {title} {\enquote {\bibinfo {title} {Two steps to
  obfuscation},}\ }\href@noop {} {\bibfield  {journal} {\bibinfo  {journal}
  {Social Networks}\ }\textbf {\bibinfo {volume} {39}},\ \bibinfo {pages}
  {12--13} (\bibinfo {year} {2014})}\BibitemShut {NoStop}%
\bibitem [{\citenamefont {Gleich}(2015)}]{gleich2015pagerank}%
  \BibitemOpen
  \bibfield  {author} {\bibinfo {author} {\bibfnamefont {D.~F.}\ \bibnamefont
  {Gleich}},\ }\bibfield  {title} {\enquote {\bibinfo {title} {Pagerank beyond
  the web},}\ }\href@noop {} {\bibfield  {journal} {\bibinfo  {journal} {SIAM
  Review}\ }\textbf {\bibinfo {volume} {57}},\ \bibinfo {pages} {321--363}
  (\bibinfo {year} {2015})}\BibitemShut {NoStop}%
\bibitem [{\citenamefont {Zhou}, \citenamefont {L{\"u}},\ and\ \citenamefont
  {Li}(2012)}]{zhou2012quantifying}%
  \BibitemOpen
  \bibfield  {author} {\bibinfo {author} {\bibfnamefont {Y.-B.}\ \bibnamefont
  {Zhou}}, \bibinfo {author} {\bibfnamefont {L.}~\bibnamefont {L{\"u}}}, \ and\
  \bibinfo {author} {\bibfnamefont {M.}~\bibnamefont {Li}},\ }\bibfield
  {title} {\enquote {\bibinfo {title} {Quantifying the influence of scientists
  and their publications: distinguishing between prestige and popularity},}\
  }\href@noop {} {\bibfield  {journal} {\bibinfo  {journal} {New Journal of
  Physics}\ }\textbf {\bibinfo {volume} {14}},\ \bibinfo {pages} {033033}
  (\bibinfo {year} {2012})}\BibitemShut {NoStop}%
\bibitem [{\citenamefont {L{\"u}}\ \emph {et~al.}(2011)\citenamefont {L{\"u}},
  \citenamefont {Zhang}, \citenamefont {Yeung},\ and\ \citenamefont
  {Zhou}}]{lu2011leaders}%
  \BibitemOpen
  \bibfield  {author} {\bibinfo {author} {\bibfnamefont {L.}~\bibnamefont
  {L{\"u}}}, \bibinfo {author} {\bibfnamefont {Y.-C.}\ \bibnamefont {Zhang}},
  \bibinfo {author} {\bibfnamefont {C.~H.}\ \bibnamefont {Yeung}}, \ and\
  \bibinfo {author} {\bibfnamefont {T.}~\bibnamefont {Zhou}},\ }\bibfield
  {title} {\enquote {\bibinfo {title} {Leaders in social networks, the
  delicious case},}\ }\href@noop {} {\bibfield  {journal} {\bibinfo  {journal}
  {PLOS ONE}\ }\textbf {\bibinfo {volume} {6}} (\bibinfo {year}
  {2011})}\BibitemShut {NoStop}%
\bibitem [{\citenamefont {Allesina}\ and\ \citenamefont
  {Pascual}(2009)}]{allesina2009googling}%
  \BibitemOpen
  \bibfield  {author} {\bibinfo {author} {\bibfnamefont {S.}~\bibnamefont
  {Allesina}}\ and\ \bibinfo {author} {\bibfnamefont {M.}~\bibnamefont
  {Pascual}},\ }\bibfield  {title} {\enquote {\bibinfo {title} {Googling food
  webs: can an eigenvector measure species' importance for coextinctions?}}\
  }\href@noop {} {\bibfield  {journal} {\bibinfo  {journal} {PLOS Computational
  Biology}\ }\textbf {\bibinfo {volume} {5}} (\bibinfo {year}
  {2009})}\BibitemShut {NoStop}%
\bibitem [{\citenamefont {Radicchi}(2011)}]{radicchi2011best}%
  \BibitemOpen
  \bibfield  {author} {\bibinfo {author} {\bibfnamefont {F.}~\bibnamefont
  {Radicchi}},\ }\bibfield  {title} {\enquote {\bibinfo {title} {Who is the
  best player ever? a complex network analysis of the history of professional
  tennis},}\ }\href@noop {} {\bibfield  {journal} {\bibinfo  {journal} {PLOS
  ONE}\ }\textbf {\bibinfo {volume} {6}} (\bibinfo {year} {2011})}\BibitemShut
  {NoStop}%
\bibitem [{\citenamefont {Abrahao}\ \emph {et~al.}(2017)\citenamefont
  {Abrahao}, \citenamefont {Parigi}, \citenamefont {Gupta},\ and\ \citenamefont
  {Cook}}]{abrahao2017reputation}%
  \BibitemOpen
  \bibfield  {author} {\bibinfo {author} {\bibfnamefont {B.}~\bibnamefont
  {Abrahao}}, \bibinfo {author} {\bibfnamefont {P.}~\bibnamefont {Parigi}},
  \bibinfo {author} {\bibfnamefont {A.}~\bibnamefont {Gupta}}, \ and\ \bibinfo
  {author} {\bibfnamefont {K.~S.}\ \bibnamefont {Cook}},\ }\bibfield  {title}
  {\enquote {\bibinfo {title} {Reputation offsets trust judgments based on
  social biases among airbnb users},}\ }\href@noop {} {\bibfield  {journal}
  {\bibinfo  {journal} {Proceedings of the National Academy of Sciences}\
  }\textbf {\bibinfo {volume} {114}},\ \bibinfo {pages} {9848--9853} (\bibinfo
  {year} {2017})}\BibitemShut {NoStop}%
\bibitem [{\citenamefont {Smith}\ and\ \citenamefont
  {Linden}(2017)}]{smith2017two}%
  \BibitemOpen
  \bibfield  {author} {\bibinfo {author} {\bibfnamefont {B.}~\bibnamefont
  {Smith}}\ and\ \bibinfo {author} {\bibfnamefont {G.}~\bibnamefont {Linden}},\
  }\bibfield  {title} {\enquote {\bibinfo {title} {Two decades of recommender
  systems at amazon. com},}\ }\href@noop {} {\bibfield  {journal} {\bibinfo
  {journal} {IEEE Internet Computing}\ }\textbf {\bibinfo {volume} {21}},\
  \bibinfo {pages} {12--18} (\bibinfo {year} {2017})}\BibitemShut {NoStop}%
\bibitem [{\citenamefont {Xu}\ \emph {et~al.}(2020{\natexlab{a}})\citenamefont
  {Xu}, \citenamefont {Zhang}, \citenamefont {L{\"u}},\ and\ \citenamefont
  {Mariani}}]{xu2020recommending}%
  \BibitemOpen
  \bibfield  {author} {\bibinfo {author} {\bibfnamefont {S.}~\bibnamefont
  {Xu}}, \bibinfo {author} {\bibfnamefont {Q.}~\bibnamefont {Zhang}}, \bibinfo
  {author} {\bibfnamefont {L.}~\bibnamefont {L{\"u}}}, \ and\ \bibinfo {author}
  {\bibfnamefont {M.~S.}\ \bibnamefont {Mariani}},\ }\bibfield  {title}
  {\enquote {\bibinfo {title} {Recommending investors for new startups by
  integrating network diffusion and investors’ domain preference},}\
  }\href@noop {} {\bibfield  {journal} {\bibinfo  {journal} {Information
  Sciences}\ }\textbf {\bibinfo {volume} {515}},\ \bibinfo {pages} {103--115}
  (\bibinfo {year} {2020}{\natexlab{a}})}\BibitemShut {NoStop}%
\bibitem [{\citenamefont {Lanz}\ \emph {et~al.}(2019)\citenamefont {Lanz},
  \citenamefont {Goldenberg}, \citenamefont {Shapira},\ and\ \citenamefont
  {Stahl}}]{lanz2019climb}%
  \BibitemOpen
  \bibfield  {author} {\bibinfo {author} {\bibfnamefont {A.}~\bibnamefont
  {Lanz}}, \bibinfo {author} {\bibfnamefont {J.}~\bibnamefont {Goldenberg}},
  \bibinfo {author} {\bibfnamefont {D.}~\bibnamefont {Shapira}}, \ and\
  \bibinfo {author} {\bibfnamefont {F.}~\bibnamefont {Stahl}},\ }\bibfield
  {title} {\enquote {\bibinfo {title} {Climb or jump: Status-based seeding in
  user-generated content networks},}\ }\href@noop {} {\bibfield  {journal}
  {\bibinfo  {journal} {Journal of Marketing Research}\ }\textbf {\bibinfo
  {volume} {56}},\ \bibinfo {pages} {361--378} (\bibinfo {year}
  {2019})}\BibitemShut {NoStop}%
\bibitem [{\citenamefont {Koren}(2009)}]{koren2009collaborative}%
  \BibitemOpen
  \bibfield  {author} {\bibinfo {author} {\bibfnamefont {Y.}~\bibnamefont
  {Koren}},\ }\bibfield  {title} {\enquote {\bibinfo {title} {Collaborative
  filtering with temporal dynamics},}\ }in\ \href@noop {} {\emph {\bibinfo
  {booktitle} {Proceedings of the 15th ACM SIGKDD International Conference on
  Knowledge Discovery and Data Mining}}}\ (\bibinfo {year} {2009})\ pp.\
  \bibinfo {pages} {447--456}\BibitemShut {NoStop}%
\bibitem [{\citenamefont {Huang}\ \emph {et~al.}(2013)\citenamefont {Huang},
  \citenamefont {He}, \citenamefont {Gao}, \citenamefont {Deng}, \citenamefont
  {Acero},\ and\ \citenamefont {Heck}}]{huang2013learning}%
  \BibitemOpen
  \bibfield  {author} {\bibinfo {author} {\bibfnamefont {P.-S.}\ \bibnamefont
  {Huang}}, \bibinfo {author} {\bibfnamefont {X.}~\bibnamefont {He}}, \bibinfo
  {author} {\bibfnamefont {J.}~\bibnamefont {Gao}}, \bibinfo {author}
  {\bibfnamefont {L.}~\bibnamefont {Deng}}, \bibinfo {author} {\bibfnamefont
  {A.}~\bibnamefont {Acero}}, \ and\ \bibinfo {author} {\bibfnamefont
  {L.}~\bibnamefont {Heck}},\ }\bibfield  {title} {\enquote {\bibinfo {title}
  {Learning deep structured semantic models for web search using clickthrough
  data},}\ }in\ \href@noop {} {\emph {\bibinfo {booktitle} {Proceedings of the
  22nd ACM International Conference on Information \& Knowledge Management}}}\
  (\bibinfo {year} {2013})\ pp.\ \bibinfo {pages} {2333--2338}\BibitemShut
  {NoStop}%
\bibitem [{\citenamefont {Kong}, \citenamefont {Sarshar},\ and\ \citenamefont
  {Roychowdhury}(2008)}]{kong2008experience}%
  \BibitemOpen
  \bibfield  {author} {\bibinfo {author} {\bibfnamefont {J.~S.}\ \bibnamefont
  {Kong}}, \bibinfo {author} {\bibfnamefont {N.}~\bibnamefont {Sarshar}}, \
  and\ \bibinfo {author} {\bibfnamefont {V.~P.}\ \bibnamefont {Roychowdhury}},\
  }\bibfield  {title} {\enquote {\bibinfo {title} {Experience versus talent
  shapes the structure of the web},}\ }\href@noop {} {\bibfield  {journal}
  {\bibinfo  {journal} {Proceedings of the National Academy of Sciences}\
  }\textbf {\bibinfo {volume} {105}},\ \bibinfo {pages} {13724--13729}
  (\bibinfo {year} {2008})}\BibitemShut {NoStop}%
\bibitem [{\citenamefont {L{\"u}}\ \emph
  {et~al.}(2016{\natexlab{a}})\citenamefont {L{\"u}}, \citenamefont {Chen},
  \citenamefont {Ren}, \citenamefont {Zhang}, \citenamefont {Zhang},\ and\
  \citenamefont {Zhou}}]{lu2016vital}%
  \BibitemOpen
  \bibfield  {author} {\bibinfo {author} {\bibfnamefont {L.}~\bibnamefont
  {L{\"u}}}, \bibinfo {author} {\bibfnamefont {D.}~\bibnamefont {Chen}},
  \bibinfo {author} {\bibfnamefont {X.-L.}\ \bibnamefont {Ren}}, \bibinfo
  {author} {\bibfnamefont {Q.-M.}\ \bibnamefont {Zhang}}, \bibinfo {author}
  {\bibfnamefont {Y.-C.}\ \bibnamefont {Zhang}}, \ and\ \bibinfo {author}
  {\bibfnamefont {T.}~\bibnamefont {Zhou}},\ }\bibfield  {title} {\enquote
  {\bibinfo {title} {Vital nodes identification in complex networks},}\
  }\href@noop {} {\bibfield  {journal} {\bibinfo  {journal} {Physics Reports}\
  }\textbf {\bibinfo {volume} {650}},\ \bibinfo {pages} {1--63} (\bibinfo
  {year} {2016}{\natexlab{a}})}\BibitemShut {NoStop}%
\bibitem [{\citenamefont {Liao}\ \emph {et~al.}(2017)\citenamefont {Liao},
  \citenamefont {Mariani}, \citenamefont {Medo}, \citenamefont {Zhang},\ and\
  \citenamefont {Zhou}}]{liao2017ranking}%
  \BibitemOpen
  \bibfield  {author} {\bibinfo {author} {\bibfnamefont {H.}~\bibnamefont
  {Liao}}, \bibinfo {author} {\bibfnamefont {M.~S.}\ \bibnamefont {Mariani}},
  \bibinfo {author} {\bibfnamefont {M.}~\bibnamefont {Medo}}, \bibinfo {author}
  {\bibfnamefont {Y.-C.}\ \bibnamefont {Zhang}}, \ and\ \bibinfo {author}
  {\bibfnamefont {M.-Y.}\ \bibnamefont {Zhou}},\ }\bibfield  {title} {\enquote
  {\bibinfo {title} {Ranking in evolving complex networks},}\ }\href@noop {}
  {\bibfield  {journal} {\bibinfo  {journal} {Physics Reports}\ }\textbf
  {\bibinfo {volume} {689}},\ \bibinfo {pages} {1--54} (\bibinfo {year}
  {2017})}\BibitemShut {NoStop}%
\bibitem [{\citenamefont {Morone}\ and\ \citenamefont
  {Makse}(2015)}]{morone2015influence}%
  \BibitemOpen
  \bibfield  {author} {\bibinfo {author} {\bibfnamefont {F.}~\bibnamefont
  {Morone}}\ and\ \bibinfo {author} {\bibfnamefont {H.~A.}\ \bibnamefont
  {Makse}},\ }\bibfield  {title} {\enquote {\bibinfo {title} {Influence
  maximization in complex networks through optimal percolation},}\ }\href@noop
  {} {\bibfield  {journal} {\bibinfo  {journal} {Nature}\ }\textbf {\bibinfo
  {volume} {524}},\ \bibinfo {pages} {65--68} (\bibinfo {year}
  {2015})}\BibitemShut {NoStop}%
\bibitem [{\citenamefont {Radicchi}\ and\ \citenamefont
  {Castellano}(2016)}]{radicchi2016leveraging}%
  \BibitemOpen
  \bibfield  {author} {\bibinfo {author} {\bibfnamefont {F.}~\bibnamefont
  {Radicchi}}\ and\ \bibinfo {author} {\bibfnamefont {C.}~\bibnamefont
  {Castellano}},\ }\bibfield  {title} {\enquote {\bibinfo {title} {Leveraging
  percolation theory to single out influential spreaders in networks},}\
  }\href@noop {} {\bibfield  {journal} {\bibinfo  {journal} {Physical Review
  E}\ }\textbf {\bibinfo {volume} {93}},\ \bibinfo {pages} {062314} (\bibinfo
  {year} {2016})}\BibitemShut {NoStop}%
\bibitem [{\citenamefont {Medo}, \citenamefont {Cimini},\ and\ \citenamefont
  {Gualdi}(2011)}]{medo2011temporal}%
  \BibitemOpen
  \bibfield  {author} {\bibinfo {author} {\bibfnamefont {M.}~\bibnamefont
  {Medo}}, \bibinfo {author} {\bibfnamefont {G.}~\bibnamefont {Cimini}}, \ and\
  \bibinfo {author} {\bibfnamefont {S.}~\bibnamefont {Gualdi}},\ }\bibfield
  {title} {\enquote {\bibinfo {title} {Temporal effects in the growth of
  networks},}\ }\href@noop {} {\bibfield  {journal} {\bibinfo  {journal}
  {Physical Review Letters}\ }\textbf {\bibinfo {volume} {107}},\ \bibinfo
  {pages} {238701} (\bibinfo {year} {2011})}\BibitemShut {NoStop}%
\bibitem [{\citenamefont {Miller}\ and\ \citenamefont
  {Page}(2009)}]{miller2009complex}%
  \BibitemOpen
  \bibfield  {author} {\bibinfo {author} {\bibfnamefont {J.~H.}\ \bibnamefont
  {Miller}}\ and\ \bibinfo {author} {\bibfnamefont {S.~E.}\ \bibnamefont
  {Page}},\ }\href@noop {} {\emph {\bibinfo {title} {Complex adaptive systems:
  An introduction to computational models of social life}}}\ (\bibinfo
  {publisher} {Princeton University Press},\ \bibinfo {year}
  {2009})\BibitemShut {NoStop}%
\bibitem [{\citenamefont {L{\"u}}\ \emph
  {et~al.}(2016{\natexlab{b}})\citenamefont {L{\"u}}, \citenamefont {Zhou},
  \citenamefont {Zhang},\ and\ \citenamefont {Stanley}}]{lu2016h}%
  \BibitemOpen
  \bibfield  {author} {\bibinfo {author} {\bibfnamefont {L.}~\bibnamefont
  {L{\"u}}}, \bibinfo {author} {\bibfnamefont {T.}~\bibnamefont {Zhou}},
  \bibinfo {author} {\bibfnamefont {Q.-M.}\ \bibnamefont {Zhang}}, \ and\
  \bibinfo {author} {\bibfnamefont {H.~E.}\ \bibnamefont {Stanley}},\
  }\bibfield  {title} {\enquote {\bibinfo {title} {The h-index of a network
  node and its relation to degree and coreness},}\ }\href@noop {} {\bibfield
  {journal} {\bibinfo  {journal} {Nature Communications}\ }\textbf {\bibinfo
  {volume} {7}},\ \bibinfo {pages} {10168} (\bibinfo {year}
  {2016}{\natexlab{b}})}\BibitemShut {NoStop}%
\bibitem [{\citenamefont {Chen}\ \emph {et~al.}(2012)\citenamefont {Chen},
  \citenamefont {L{\"u}}, \citenamefont {Shang}, \citenamefont {Zhang},\ and\
  \citenamefont {Zhou}}]{chen2012identifying}%
  \BibitemOpen
  \bibfield  {author} {\bibinfo {author} {\bibfnamefont {D.}~\bibnamefont
  {Chen}}, \bibinfo {author} {\bibfnamefont {L.}~\bibnamefont {L{\"u}}},
  \bibinfo {author} {\bibfnamefont {M.-S.}\ \bibnamefont {Shang}}, \bibinfo
  {author} {\bibfnamefont {Y.-C.}\ \bibnamefont {Zhang}}, \ and\ \bibinfo
  {author} {\bibfnamefont {T.}~\bibnamefont {Zhou}},\ }\bibfield  {title}
  {\enquote {\bibinfo {title} {Identifying influential nodes in complex
  networks},}\ }\href@noop {} {\bibfield  {journal} {\bibinfo  {journal}
  {Physica A: Statistical Mechanics and Its Applications}\ }\textbf {\bibinfo
  {volume} {391}},\ \bibinfo {pages} {1777--1787} (\bibinfo {year}
  {2012})}\BibitemShut {NoStop}%
\bibitem [{\citenamefont {Iannelli}, \citenamefont {Mariani},\ and\
  \citenamefont {Sokolov}(2018)}]{iannelli2018influencers}%
  \BibitemOpen
  \bibfield  {author} {\bibinfo {author} {\bibfnamefont {F.}~\bibnamefont
  {Iannelli}}, \bibinfo {author} {\bibfnamefont {M.~S.}\ \bibnamefont
  {Mariani}}, \ and\ \bibinfo {author} {\bibfnamefont {I.~M.}\ \bibnamefont
  {Sokolov}},\ }\bibfield  {title} {\enquote {\bibinfo {title} {Influencers
  identification in complex networks through reaction-diffusion dynamics},}\
  }\href@noop {} {\bibfield  {journal} {\bibinfo  {journal} {Physical Review
  E}\ }\textbf {\bibinfo {volume} {98}},\ \bibinfo {pages} {062302} (\bibinfo
  {year} {2018})}\BibitemShut {NoStop}%
\bibitem [{\citenamefont {Brandes}(2001)}]{brandes2001faster}%
  \BibitemOpen
  \bibfield  {author} {\bibinfo {author} {\bibfnamefont {U.}~\bibnamefont
  {Brandes}},\ }\bibfield  {title} {\enquote {\bibinfo {title} {A faster
  algorithm for betweenness centrality},}\ }\href@noop {} {\bibfield  {journal}
  {\bibinfo  {journal} {Journal of Mathematical Sociology}\ }\textbf {\bibinfo
  {volume} {25}},\ \bibinfo {pages} {163--177} (\bibinfo {year}
  {2001})}\BibitemShut {NoStop}%
\bibitem [{\citenamefont {Bradley}\ and\ \citenamefont
  {Terry}(1952)}]{bradley1952rank}%
  \BibitemOpen
  \bibfield  {author} {\bibinfo {author} {\bibfnamefont {R.~A.}\ \bibnamefont
  {Bradley}}\ and\ \bibinfo {author} {\bibfnamefont {M.~E.}\ \bibnamefont
  {Terry}},\ }\bibfield  {title} {\enquote {\bibinfo {title} {Rank analysis of
  incomplete block designs: I. the method of paired comparisons},}\ }\href@noop
  {} {\bibfield  {journal} {\bibinfo  {journal} {Biometrika}\ }\textbf
  {\bibinfo {volume} {39}},\ \bibinfo {pages} {324--345} (\bibinfo {year}
  {1952})}\BibitemShut {NoStop}%
\bibitem [{\citenamefont {Fogel}, \citenamefont {d'Aspremont},\ and\
  \citenamefont {Vojnovic}(2016)}]{fogel2016spectral}%
  \BibitemOpen
  \bibfield  {author} {\bibinfo {author} {\bibfnamefont {F.}~\bibnamefont
  {Fogel}}, \bibinfo {author} {\bibfnamefont {A.}~\bibnamefont {d'Aspremont}},
  \ and\ \bibinfo {author} {\bibfnamefont {M.}~\bibnamefont {Vojnovic}},\
  }\bibfield  {title} {\enquote {\bibinfo {title} {Spectral ranking using
  seriation},}\ }\href@noop {} {\bibfield  {journal} {\bibinfo  {journal} {The
  Journal of Machine Learning Research}\ }\textbf {\bibinfo {volume} {17}},\
  \bibinfo {pages} {3013--3057} (\bibinfo {year} {2016})}\BibitemShut {NoStop}%
\bibitem [{\citenamefont {Cucuringu}(2016)}]{cucuringu2016sync}%
  \BibitemOpen
  \bibfield  {author} {\bibinfo {author} {\bibfnamefont {M.}~\bibnamefont
  {Cucuringu}},\ }\bibfield  {title} {\enquote {\bibinfo {title} {Sync-rank:
  Robust ranking, constrained ranking and rank aggregation via eigenvector and
  sdp synchronization},}\ }\href@noop {} {\bibfield  {journal} {\bibinfo
  {journal} {IEEE Transactions on Network Science and Engineering}\ }\textbf
  {\bibinfo {volume} {3}},\ \bibinfo {pages} {58--79} (\bibinfo {year}
  {2016})}\BibitemShut {NoStop}%
\bibitem [{\citenamefont {De~Bacco}, \citenamefont {Larremore},\ and\
  \citenamefont {Moore}(2018)}]{de2018physical}%
  \BibitemOpen
  \bibfield  {author} {\bibinfo {author} {\bibfnamefont {C.}~\bibnamefont
  {De~Bacco}}, \bibinfo {author} {\bibfnamefont {D.~B.}\ \bibnamefont
  {Larremore}}, \ and\ \bibinfo {author} {\bibfnamefont {C.}~\bibnamefont
  {Moore}},\ }\bibfield  {title} {\enquote {\bibinfo {title} {A physical model
  for efficient ranking in networks},}\ }\href@noop {} {\bibfield  {journal}
  {\bibinfo  {journal} {Science advances}\ }\textbf {\bibinfo {volume} {4}},\
  \bibinfo {pages} {eaar8260} (\bibinfo {year} {2018})}\BibitemShut {NoStop}%
\bibitem [{\citenamefont {d'Aspremont}, \citenamefont {Cucuringu},\ and\
  \citenamefont {Tyagi}(2019)}]{d2019ranking}%
  \BibitemOpen
  \bibfield  {author} {\bibinfo {author} {\bibfnamefont {A.}~\bibnamefont
  {d'Aspremont}}, \bibinfo {author} {\bibfnamefont {M.}~\bibnamefont
  {Cucuringu}}, \ and\ \bibinfo {author} {\bibfnamefont {H.}~\bibnamefont
  {Tyagi}},\ }\bibfield  {title} {\enquote {\bibinfo {title} {Ranking and
  synchronization from pairwise measurements via svd},}\ }\href@noop {}
  {\bibfield  {journal} {\bibinfo  {journal} {arXiv preprint arXiv:1906.02746}\
  } (\bibinfo {year} {2019})}\BibitemShut {NoStop}%
\bibitem [{\citenamefont {Mariani}, \citenamefont {Medo},\ and\ \citenamefont
  {Zhang}(2015)}]{mariani2015ranking}%
  \BibitemOpen
  \bibfield  {author} {\bibinfo {author} {\bibfnamefont {M.~S.}\ \bibnamefont
  {Mariani}}, \bibinfo {author} {\bibfnamefont {M.}~\bibnamefont {Medo}}, \
  and\ \bibinfo {author} {\bibfnamefont {Y.-C.}\ \bibnamefont {Zhang}},\
  }\bibfield  {title} {\enquote {\bibinfo {title} {Ranking nodes in growing
  networks: When pagerank fails},}\ }\href@noop {} {\bibfield  {journal}
  {\bibinfo  {journal} {Scientific Reports}\ }\textbf {\bibinfo {volume} {5}},\
  \bibinfo {pages} {16181} (\bibinfo {year} {2015})}\BibitemShut {NoStop}%
\bibitem [{\citenamefont {Ntoutsi}\ \emph {et~al.}(2020)\citenamefont
  {Ntoutsi}, \citenamefont {Fafalios}, \citenamefont {Gadiraju}, \citenamefont
  {Iosifidis}, \citenamefont {Nejdl}, \citenamefont {Vidal}, \citenamefont
  {Ruggieri}, \citenamefont {Turini}, \citenamefont {Papadopoulos},
  \citenamefont {Krasanakis} \emph {et~al.}}]{ntoutsi2020bias}%
  \BibitemOpen
  \bibfield  {author} {\bibinfo {author} {\bibfnamefont {E.}~\bibnamefont
  {Ntoutsi}}, \bibinfo {author} {\bibfnamefont {P.}~\bibnamefont {Fafalios}},
  \bibinfo {author} {\bibfnamefont {U.}~\bibnamefont {Gadiraju}}, \bibinfo
  {author} {\bibfnamefont {V.}~\bibnamefont {Iosifidis}}, \bibinfo {author}
  {\bibfnamefont {W.}~\bibnamefont {Nejdl}}, \bibinfo {author} {\bibfnamefont
  {M.-E.}\ \bibnamefont {Vidal}}, \bibinfo {author} {\bibfnamefont
  {S.}~\bibnamefont {Ruggieri}}, \bibinfo {author} {\bibfnamefont
  {F.}~\bibnamefont {Turini}}, \bibinfo {author} {\bibfnamefont
  {S.}~\bibnamefont {Papadopoulos}}, \bibinfo {author} {\bibfnamefont
  {E.}~\bibnamefont {Krasanakis}},  \emph {et~al.},\ }\bibfield  {title}
  {\enquote {\bibinfo {title} {Bias in data-driven ai systems--an introductory
  survey},}\ }\href@noop {} {\bibfield  {journal} {\bibinfo  {journal} {arXiv
  preprint arXiv:2001.09762}\ } (\bibinfo {year} {2020})}\BibitemShut {NoStop}%
\bibitem [{\citenamefont {Abbott}\ \emph {et~al.}(2016)\citenamefont {Abbott},
  \citenamefont {Abbott}, \citenamefont {Abbott}, \citenamefont {Abernathy},
  \citenamefont {Acernese}, \citenamefont {Ackley}, \citenamefont {Adams},
  \citenamefont {Adams}, \citenamefont {Addesso}, \citenamefont {Adhikari}
  \emph {et~al.}}]{abbott2016observation}%
  \BibitemOpen
  \bibfield  {author} {\bibinfo {author} {\bibfnamefont {B.~P.}\ \bibnamefont
  {Abbott}}, \bibinfo {author} {\bibfnamefont {R.}~\bibnamefont {Abbott}},
  \bibinfo {author} {\bibfnamefont {T.}~\bibnamefont {Abbott}}, \bibinfo
  {author} {\bibfnamefont {M.}~\bibnamefont {Abernathy}}, \bibinfo {author}
  {\bibfnamefont {F.}~\bibnamefont {Acernese}}, \bibinfo {author}
  {\bibfnamefont {K.}~\bibnamefont {Ackley}}, \bibinfo {author} {\bibfnamefont
  {C.}~\bibnamefont {Adams}}, \bibinfo {author} {\bibfnamefont
  {T.}~\bibnamefont {Adams}}, \bibinfo {author} {\bibfnamefont
  {P.}~\bibnamefont {Addesso}}, \bibinfo {author} {\bibfnamefont
  {R.}~\bibnamefont {Adhikari}},  \emph {et~al.},\ }\bibfield  {title}
  {\enquote {\bibinfo {title} {Observation of gravitational waves from a binary
  black hole merger},}\ }\href@noop {} {\bibfield  {journal} {\bibinfo
  {journal} {Physical Review Letters}\ }\textbf {\bibinfo {volume} {116}},\
  \bibinfo {pages} {061102} (\bibinfo {year} {2016})}\BibitemShut {NoStop}%
\bibitem [{\citenamefont {Fortunato}\ \emph {et~al.}(2006)\citenamefont
  {Fortunato}, \citenamefont {Bogu{\~n}{\'a}}, \citenamefont {Flammini},\ and\
  \citenamefont {Menczer}}]{fortunato2006approximating}%
  \BibitemOpen
  \bibfield  {author} {\bibinfo {author} {\bibfnamefont {S.}~\bibnamefont
  {Fortunato}}, \bibinfo {author} {\bibfnamefont {M.}~\bibnamefont
  {Bogu{\~n}{\'a}}}, \bibinfo {author} {\bibfnamefont {A.}~\bibnamefont
  {Flammini}}, \ and\ \bibinfo {author} {\bibfnamefont {F.}~\bibnamefont
  {Menczer}},\ }\bibfield  {title} {\enquote {\bibinfo {title} {Approximating
  pagerank from in-degree},}\ }in\ \href@noop {} {\emph {\bibinfo {booktitle}
  {International Workshop on Algorithms and Models for the Web-Graph}}}\
  (\bibinfo {organization} {Springer},\ \bibinfo {year} {2006})\ pp.\ \bibinfo
  {pages} {59--71}\BibitemShut {NoStop}%
\bibitem [{\citenamefont {Mariani}, \citenamefont {Medo},\ and\ \citenamefont
  {Zhang}(2016)}]{mariani2016identification}%
  \BibitemOpen
  \bibfield  {author} {\bibinfo {author} {\bibfnamefont {M.~S.}\ \bibnamefont
  {Mariani}}, \bibinfo {author} {\bibfnamefont {M.}~\bibnamefont {Medo}}, \
  and\ \bibinfo {author} {\bibfnamefont {Y.-C.}\ \bibnamefont {Zhang}},\
  }\bibfield  {title} {\enquote {\bibinfo {title} {Identification of milestone
  papers through time-balanced network centrality},}\ }\href@noop {} {\bibfield
   {journal} {\bibinfo  {journal} {Journal of Informetrics}\ }\textbf {\bibinfo
  {volume} {10}},\ \bibinfo {pages} {1207--1223} (\bibinfo {year}
  {2016})}\BibitemShut {NoStop}%
\bibitem [{\citenamefont {Xu}\ \emph {et~al.}(2020{\natexlab{b}})\citenamefont
  {Xu}, \citenamefont {Mariani}, \citenamefont {L{\"u}},\ and\ \citenamefont
  {Medo}}]{xu2020unbiased}%
  \BibitemOpen
  \bibfield  {author} {\bibinfo {author} {\bibfnamefont {S.}~\bibnamefont
  {Xu}}, \bibinfo {author} {\bibfnamefont {M.~S.}\ \bibnamefont {Mariani}},
  \bibinfo {author} {\bibfnamefont {L.}~\bibnamefont {L{\"u}}}, \ and\ \bibinfo
  {author} {\bibfnamefont {M.}~\bibnamefont {Medo}},\ }\bibfield  {title}
  {\enquote {\bibinfo {title} {Unbiased evaluation of ranking metrics reveals
  consistent performance in science and technology citation data},}\
  }\href@noop {} {\bibfield  {journal} {\bibinfo  {journal} {Journal of
  Informetrics}\ }\textbf {\bibinfo {volume} {14}},\ \bibinfo {pages} {101005}
  (\bibinfo {year} {2020}{\natexlab{b}})}\BibitemShut {NoStop}%
\bibitem [{\citenamefont {Mariani}, \citenamefont {Medo},\ and\ \citenamefont
  {Lafond}(2019)}]{mariani2019early}%
  \BibitemOpen
  \bibfield  {author} {\bibinfo {author} {\bibfnamefont {M.~S.}\ \bibnamefont
  {Mariani}}, \bibinfo {author} {\bibfnamefont {M.}~\bibnamefont {Medo}}, \
  and\ \bibinfo {author} {\bibfnamefont {F.}~\bibnamefont {Lafond}},\
  }\bibfield  {title} {\enquote {\bibinfo {title} {Early identification of
  important patents: Design and validation of citation network metrics},}\
  }\href@noop {} {\bibfield  {journal} {\bibinfo  {journal} {Technological
  Forecasting and Social Change}\ }\textbf {\bibinfo {volume} {146}},\ \bibinfo
  {pages} {644--654} (\bibinfo {year} {2019})}\BibitemShut {NoStop}%
\bibitem [{\citenamefont {Ren}\ \emph {et~al.}(2018)\citenamefont {Ren},
  \citenamefont {Mariani}, \citenamefont {Zhang},\ and\ \citenamefont
  {Medo}}]{ren2018randomizing}%
  \BibitemOpen
  \bibfield  {author} {\bibinfo {author} {\bibfnamefont {Z.-M.}\ \bibnamefont
  {Ren}}, \bibinfo {author} {\bibfnamefont {M.~S.}\ \bibnamefont {Mariani}},
  \bibinfo {author} {\bibfnamefont {Y.-C.}\ \bibnamefont {Zhang}}, \ and\
  \bibinfo {author} {\bibfnamefont {M.}~\bibnamefont {Medo}},\ }\bibfield
  {title} {\enquote {\bibinfo {title} {Randomizing growing networks with a
  time-respecting null model},}\ }\href@noop {} {\bibfield  {journal} {\bibinfo
   {journal} {Physical Review E}\ }\textbf {\bibinfo {volume} {97}},\ \bibinfo
  {pages} {052311} (\bibinfo {year} {2018})}\BibitemShut {NoStop}%
\bibitem [{\citenamefont {Waltman}(2016)}]{waltman2016review}%
  \BibitemOpen
  \bibfield  {author} {\bibinfo {author} {\bibfnamefont {L.}~\bibnamefont
  {Waltman}},\ }\bibfield  {title} {\enquote {\bibinfo {title} {A review of the
  literature on citation impact indicators},}\ }\href@noop {} {\bibfield
  {journal} {\bibinfo  {journal} {Journal of Informetrics}\ }\textbf {\bibinfo
  {volume} {10}},\ \bibinfo {pages} {365--391} (\bibinfo {year}
  {2016})}\BibitemShut {NoStop}%
\bibitem [{\citenamefont {Battiston}\ \emph {et~al.}(2019)\citenamefont
  {Battiston}, \citenamefont {Musciotto}, \citenamefont {Wang}, \citenamefont
  {Barab{\'a}si}, \citenamefont {Szell},\ and\ \citenamefont
  {Sinatra}}]{battiston2019taking}%
  \BibitemOpen
  \bibfield  {author} {\bibinfo {author} {\bibfnamefont {F.}~\bibnamefont
  {Battiston}}, \bibinfo {author} {\bibfnamefont {F.}~\bibnamefont
  {Musciotto}}, \bibinfo {author} {\bibfnamefont {D.}~\bibnamefont {Wang}},
  \bibinfo {author} {\bibfnamefont {A.-L.}\ \bibnamefont {Barab{\'a}si}},
  \bibinfo {author} {\bibfnamefont {M.}~\bibnamefont {Szell}}, \ and\ \bibinfo
  {author} {\bibfnamefont {R.}~\bibnamefont {Sinatra}},\ }\bibfield  {title}
  {\enquote {\bibinfo {title} {Taking census of physics},}\ }\href@noop {}
  {\bibfield  {journal} {\bibinfo  {journal} {Nature Reviews Physics}\ }\textbf
  {\bibinfo {volume} {1}},\ \bibinfo {pages} {89--97} (\bibinfo {year}
  {2019})}\BibitemShut {NoStop}%
\bibitem [{\citenamefont {Vaccario}\ \emph {et~al.}(2017)\citenamefont
  {Vaccario}, \citenamefont {Medo}, \citenamefont {Wider},\ and\ \citenamefont
  {Mariani}}]{vaccario2017quantifying}%
  \BibitemOpen
  \bibfield  {author} {\bibinfo {author} {\bibfnamefont {G.}~\bibnamefont
  {Vaccario}}, \bibinfo {author} {\bibfnamefont {M.}~\bibnamefont {Medo}},
  \bibinfo {author} {\bibfnamefont {N.}~\bibnamefont {Wider}}, \ and\ \bibinfo
  {author} {\bibfnamefont {M.~S.}\ \bibnamefont {Mariani}},\ }\bibfield
  {title} {\enquote {\bibinfo {title} {Quantifying and suppressing ranking bias
  in a large citation network},}\ }\href@noop {} {\bibfield  {journal}
  {\bibinfo  {journal} {Journal of informetrics}\ }\textbf {\bibinfo {volume}
  {11}},\ \bibinfo {pages} {766--782} (\bibinfo {year} {2017})}\BibitemShut
  {NoStop}%
\bibitem [{\citenamefont {Dunaiski}, \citenamefont {Geldenhuys},\ and\
  \citenamefont {Visser}(2019{\natexlab{a}})}]{dunaiski2019globalised}%
  \BibitemOpen
  \bibfield  {author} {\bibinfo {author} {\bibfnamefont {M.}~\bibnamefont
  {Dunaiski}}, \bibinfo {author} {\bibfnamefont {J.}~\bibnamefont
  {Geldenhuys}}, \ and\ \bibinfo {author} {\bibfnamefont {W.}~\bibnamefont
  {Visser}},\ }\bibfield  {title} {\enquote {\bibinfo {title} {Globalised vs
  averaged: Bias and ranking performance on the author level},}\ }\href@noop {}
  {\bibfield  {journal} {\bibinfo  {journal} {Journal of Informetrics}\
  }\textbf {\bibinfo {volume} {13}},\ \bibinfo {pages} {299--313} (\bibinfo
  {year} {2019}{\natexlab{a}})}\BibitemShut {NoStop}%
\bibitem [{\citenamefont {Petersen}\ \emph {et~al.}(2014)\citenamefont
  {Petersen}, \citenamefont {Fortunato}, \citenamefont {Pan}, \citenamefont
  {Kaski}, \citenamefont {Penner}, \citenamefont {Rungi}, \citenamefont
  {Riccaboni}, \citenamefont {Stanley},\ and\ \citenamefont
  {Pammolli}}]{petersen2014reputation}%
  \BibitemOpen
  \bibfield  {author} {\bibinfo {author} {\bibfnamefont {A.~M.}\ \bibnamefont
  {Petersen}}, \bibinfo {author} {\bibfnamefont {S.}~\bibnamefont {Fortunato}},
  \bibinfo {author} {\bibfnamefont {R.~K.}\ \bibnamefont {Pan}}, \bibinfo
  {author} {\bibfnamefont {K.}~\bibnamefont {Kaski}}, \bibinfo {author}
  {\bibfnamefont {O.}~\bibnamefont {Penner}}, \bibinfo {author} {\bibfnamefont
  {A.}~\bibnamefont {Rungi}}, \bibinfo {author} {\bibfnamefont
  {M.}~\bibnamefont {Riccaboni}}, \bibinfo {author} {\bibfnamefont {H.~E.}\
  \bibnamefont {Stanley}}, \ and\ \bibinfo {author} {\bibfnamefont
  {F.}~\bibnamefont {Pammolli}},\ }\bibfield  {title} {\enquote {\bibinfo
  {title} {Reputation and impact in academic careers},}\ }\href@noop {}
  {\bibfield  {journal} {\bibinfo  {journal} {Proceedings of the National
  Academy of Sciences}\ }\textbf {\bibinfo {volume} {111}},\ \bibinfo {pages}
  {15316--15321} (\bibinfo {year} {2014})}\BibitemShut {NoStop}%
\bibitem [{\citenamefont {Sarig{\"o}l}\ \emph {et~al.}(2014)\citenamefont
  {Sarig{\"o}l}, \citenamefont {Pfitzner}, \citenamefont {Scholtes},
  \citenamefont {Garas},\ and\ \citenamefont
  {Schweitzer}}]{sarigol2014predicting}%
  \BibitemOpen
  \bibfield  {author} {\bibinfo {author} {\bibfnamefont {E.}~\bibnamefont
  {Sarig{\"o}l}}, \bibinfo {author} {\bibfnamefont {R.}~\bibnamefont
  {Pfitzner}}, \bibinfo {author} {\bibfnamefont {I.}~\bibnamefont {Scholtes}},
  \bibinfo {author} {\bibfnamefont {A.}~\bibnamefont {Garas}}, \ and\ \bibinfo
  {author} {\bibfnamefont {F.}~\bibnamefont {Schweitzer}},\ }\bibfield  {title}
  {\enquote {\bibinfo {title} {Predicting scientific success based on
  coauthorship networks},}\ }\href@noop {} {\bibfield  {journal} {\bibinfo
  {journal} {EPJ Data Science}\ }\textbf {\bibinfo {volume} {3}},\ \bibinfo
  {pages} {9} (\bibinfo {year} {2014})}\BibitemShut {NoStop}%
\bibitem [{\citenamefont {Li}\ \emph {et~al.}(2019)\citenamefont {Li},
  \citenamefont {Aste}, \citenamefont {Caccioli},\ and\ \citenamefont
  {Livan}}]{li2019reciprocity}%
  \BibitemOpen
  \bibfield  {author} {\bibinfo {author} {\bibfnamefont {W.}~\bibnamefont
  {Li}}, \bibinfo {author} {\bibfnamefont {T.}~\bibnamefont {Aste}}, \bibinfo
  {author} {\bibfnamefont {F.}~\bibnamefont {Caccioli}}, \ and\ \bibinfo
  {author} {\bibfnamefont {G.}~\bibnamefont {Livan}},\ }\bibfield  {title}
  {\enquote {\bibinfo {title} {Reciprocity and impact in academic careers},}\
  }\href@noop {} {\bibfield  {journal} {\bibinfo  {journal} {EPJ Data Science}\
  }\textbf {\bibinfo {volume} {8}},\ \bibinfo {pages} {20} (\bibinfo {year}
  {2019})}\BibitemShut {NoStop}%
\bibitem [{\citenamefont {Kitsak}\ \emph {et~al.}(2010)\citenamefont {Kitsak},
  \citenamefont {Gallos}, \citenamefont {Havlin}, \citenamefont {Liljeros},
  \citenamefont {Muchnik}, \citenamefont {Stanley},\ and\ \citenamefont
  {Makse}}]{kitsak2010identification}%
  \BibitemOpen
  \bibfield  {author} {\bibinfo {author} {\bibfnamefont {M.}~\bibnamefont
  {Kitsak}}, \bibinfo {author} {\bibfnamefont {L.~K.}\ \bibnamefont {Gallos}},
  \bibinfo {author} {\bibfnamefont {S.}~\bibnamefont {Havlin}}, \bibinfo
  {author} {\bibfnamefont {F.}~\bibnamefont {Liljeros}}, \bibinfo {author}
  {\bibfnamefont {L.}~\bibnamefont {Muchnik}}, \bibinfo {author} {\bibfnamefont
  {H.~E.}\ \bibnamefont {Stanley}}, \ and\ \bibinfo {author} {\bibfnamefont
  {H.~A.}\ \bibnamefont {Makse}},\ }\bibfield  {title} {\enquote {\bibinfo
  {title} {Identification of influential spreaders in complex networks},}\
  }\href@noop {} {\bibfield  {journal} {\bibinfo  {journal} {Nature Physics}\
  }\textbf {\bibinfo {volume} {6}},\ \bibinfo {pages} {888--893} (\bibinfo
  {year} {2010})}\BibitemShut {NoStop}%
\bibitem [{\citenamefont {Fan}\ \emph {et~al.}(2020)\citenamefont {Fan},
  \citenamefont {L{\"u}}, \citenamefont {Shi},\ and\ \citenamefont
  {Zhou}}]{fan2020characterizing}%
  \BibitemOpen
  \bibfield  {author} {\bibinfo {author} {\bibfnamefont {T.}~\bibnamefont
  {Fan}}, \bibinfo {author} {\bibfnamefont {L.}~\bibnamefont {L{\"u}}},
  \bibinfo {author} {\bibfnamefont {D.}~\bibnamefont {Shi}}, \ and\ \bibinfo
  {author} {\bibfnamefont {T.}~\bibnamefont {Zhou}},\ }\bibfield  {title}
  {\enquote {\bibinfo {title} {Characterizing cycle structure in complex
  networks},}\ }\href@noop {} {\bibfield  {journal} {\bibinfo  {journal} {arXiv
  preprint arXiv:2001.08541}\ } (\bibinfo {year} {2020})}\BibitemShut {NoStop}%
\bibitem [{\citenamefont {Dunaiski}, \citenamefont {Geldenhuys},\ and\
  \citenamefont {Visser}(2018)}]{dunaiski2018author}%
  \BibitemOpen
  \bibfield  {author} {\bibinfo {author} {\bibfnamefont {M.}~\bibnamefont
  {Dunaiski}}, \bibinfo {author} {\bibfnamefont {J.}~\bibnamefont
  {Geldenhuys}}, \ and\ \bibinfo {author} {\bibfnamefont {W.}~\bibnamefont
  {Visser}},\ }\bibfield  {title} {\enquote {\bibinfo {title} {Author ranking
  evaluation at scale},}\ }\href@noop {} {\bibfield  {journal} {\bibinfo
  {journal} {Journal of Informetrics}\ }\textbf {\bibinfo {volume} {12}},\
  \bibinfo {pages} {679--702} (\bibinfo {year} {2018})}\BibitemShut {NoStop}%
\bibitem [{\citenamefont {Zhou}, \citenamefont {L{\"u}},\ and\ \citenamefont
  {Mariani}(2019)}]{zhou2019fast}%
  \BibitemOpen
  \bibfield  {author} {\bibinfo {author} {\bibfnamefont {F.}~\bibnamefont
  {Zhou}}, \bibinfo {author} {\bibfnamefont {L.}~\bibnamefont {L{\"u}}}, \ and\
  \bibinfo {author} {\bibfnamefont {M.~S.}\ \bibnamefont {Mariani}},\
  }\bibfield  {title} {\enquote {\bibinfo {title} {Fast influencers in complex
  networks},}\ }\href@noop {} {\bibfield  {journal} {\bibinfo  {journal}
  {Communications in Nonlinear Science and Numerical Simulation}\ }\textbf
  {\bibinfo {volume} {74}},\ \bibinfo {pages} {69--83} (\bibinfo {year}
  {2019})}\BibitemShut {NoStop}%
\bibitem [{\citenamefont {Dunaiski}, \citenamefont {Geldenhuys},\ and\
  \citenamefont {Visser}(2019{\natexlab{b}})}]{dunaiski2019interplay}%
  \BibitemOpen
  \bibfield  {author} {\bibinfo {author} {\bibfnamefont {M.}~\bibnamefont
  {Dunaiski}}, \bibinfo {author} {\bibfnamefont {J.}~\bibnamefont
  {Geldenhuys}}, \ and\ \bibinfo {author} {\bibfnamefont {W.}~\bibnamefont
  {Visser}},\ }\bibfield  {title} {\enquote {\bibinfo {title} {On the interplay
  between normalisation, bias, and performance of paper impact metrics},}\
  }\href@noop {} {\bibfield  {journal} {\bibinfo  {journal} {Journal of
  Informetrics}\ }\textbf {\bibinfo {volume} {13}},\ \bibinfo {pages}
  {270--290} (\bibinfo {year} {2019}{\natexlab{b}})}\BibitemShut {NoStop}%
\bibitem [{\citenamefont {Rijcke}\ \emph {et~al.}(2016)\citenamefont {Rijcke},
  \citenamefont {Wouters}, \citenamefont {Rushforth}, \citenamefont
  {Franssen},\ and\ \citenamefont {Hammarfelt}}]{rijcke2016evaluation}%
  \BibitemOpen
  \bibfield  {author} {\bibinfo {author} {\bibfnamefont {S.~d.}\ \bibnamefont
  {Rijcke}}, \bibinfo {author} {\bibfnamefont {P.~F.}\ \bibnamefont {Wouters}},
  \bibinfo {author} {\bibfnamefont {A.~D.}\ \bibnamefont {Rushforth}}, \bibinfo
  {author} {\bibfnamefont {T.~P.}\ \bibnamefont {Franssen}}, \ and\ \bibinfo
  {author} {\bibfnamefont {B.}~\bibnamefont {Hammarfelt}},\ }\bibfield  {title}
  {\enquote {\bibinfo {title} {Evaluation practices and effects of indicator
  use—a literature review},}\ }\href@noop {} {\bibfield  {journal} {\bibinfo
  {journal} {Research Evaluation}\ }\textbf {\bibinfo {volume} {25}},\ \bibinfo
  {pages} {161--169} (\bibinfo {year} {2016})}\BibitemShut {NoStop}%
\bibitem [{\citenamefont {Zhang}\ \emph {et~al.}(2019)\citenamefont {Zhang},
  \citenamefont {Medo}, \citenamefont {L{\"u}},\ and\ \citenamefont
  {Mariani}}]{zhang2019long}%
  \BibitemOpen
  \bibfield  {author} {\bibinfo {author} {\bibfnamefont {S.}~\bibnamefont
  {Zhang}}, \bibinfo {author} {\bibfnamefont {M.}~\bibnamefont {Medo}},
  \bibinfo {author} {\bibfnamefont {L.}~\bibnamefont {L{\"u}}}, \ and\ \bibinfo
  {author} {\bibfnamefont {M.~S.}\ \bibnamefont {Mariani}},\ }\bibfield
  {title} {\enquote {\bibinfo {title} {The long-term impact of ranking
  algorithms in growing networks},}\ }\href@noop {} {\bibfield  {journal}
  {\bibinfo  {journal} {Information Sciences}\ }\textbf {\bibinfo {volume}
  {488}},\ \bibinfo {pages} {257--271} (\bibinfo {year} {2019})}\BibitemShut
  {NoStop}%
\bibitem [{\citenamefont {Salganik}, \citenamefont {Dodds},\ and\ \citenamefont
  {Watts}(2006)}]{salganik2006experimental}%
  \BibitemOpen
  \bibfield  {author} {\bibinfo {author} {\bibfnamefont {M.~J.}\ \bibnamefont
  {Salganik}}, \bibinfo {author} {\bibfnamefont {P.~S.}\ \bibnamefont {Dodds}},
  \ and\ \bibinfo {author} {\bibfnamefont {D.~J.}\ \bibnamefont {Watts}},\
  }\bibfield  {title} {\enquote {\bibinfo {title} {Experimental study of
  inequality and unpredictability in an artificial cultural market},}\
  }\href@noop {} {\bibfield  {journal} {\bibinfo  {journal} {Science}\ }\textbf
  {\bibinfo {volume} {311}},\ \bibinfo {pages} {854--856} (\bibinfo {year}
  {2006})}\BibitemShut {NoStop}%
\bibitem [{\citenamefont {Bardoscia}\ \emph {et~al.}(2013)\citenamefont
  {Bardoscia}, \citenamefont {De~Luca}, \citenamefont {Livan}, \citenamefont
  {Marsili},\ and\ \citenamefont {Tessone}}]{bardoscia2013social}%
  \BibitemOpen
  \bibfield  {author} {\bibinfo {author} {\bibfnamefont {M.}~\bibnamefont
  {Bardoscia}}, \bibinfo {author} {\bibfnamefont {G.}~\bibnamefont {De~Luca}},
  \bibinfo {author} {\bibfnamefont {G.}~\bibnamefont {Livan}}, \bibinfo
  {author} {\bibfnamefont {M.}~\bibnamefont {Marsili}}, \ and\ \bibinfo
  {author} {\bibfnamefont {C.~J.}\ \bibnamefont {Tessone}},\ }\bibfield
  {title} {\enquote {\bibinfo {title} {The social climbing game},}\ }\href@noop
  {} {\bibfield  {journal} {\bibinfo  {journal} {Journal of Statistical
  Physics}\ }\textbf {\bibinfo {volume} {151}},\ \bibinfo {pages} {440--457}
  (\bibinfo {year} {2013})}\BibitemShut {NoStop}%
\bibitem [{\citenamefont {Zeng}\ \emph {et~al.}(2015)\citenamefont {Zeng},
  \citenamefont {Yeung}, \citenamefont {Medo},\ and\ \citenamefont
  {Zhang}}]{zeng2015modeling}%
  \BibitemOpen
  \bibfield  {author} {\bibinfo {author} {\bibfnamefont {A.}~\bibnamefont
  {Zeng}}, \bibinfo {author} {\bibfnamefont {C.~H.}\ \bibnamefont {Yeung}},
  \bibinfo {author} {\bibfnamefont {M.}~\bibnamefont {Medo}}, \ and\ \bibinfo
  {author} {\bibfnamefont {Y.-C.}\ \bibnamefont {Zhang}},\ }\bibfield  {title}
  {\enquote {\bibinfo {title} {Modeling mutual feedback between users and
  recommender systems},}\ }\href@noop {} {\bibfield  {journal} {\bibinfo
  {journal} {Journal of Statistical Mechanics: Theory and Experiment}\ }\textbf
  {\bibinfo {volume} {2015}},\ \bibinfo {pages} {P07020} (\bibinfo {year}
  {2015})}\BibitemShut {NoStop}%
\bibitem [{\citenamefont {Livan}(2019)}]{livan2019don}%
  \BibitemOpen
  \bibfield  {author} {\bibinfo {author} {\bibfnamefont {G.}~\bibnamefont
  {Livan}},\ }\bibfield  {title} {\enquote {\bibinfo {title} {Don’t follow
  the leader: how ranking performance reduces meritocracy},}\ }\href@noop {}
  {\bibfield  {journal} {\bibinfo  {journal} {Royal Society Open Science}\
  }\textbf {\bibinfo {volume} {6}},\ \bibinfo {pages} {191255} (\bibinfo {year}
  {2019})}\BibitemShut {NoStop}%
\bibitem [{\citenamefont {D'Amour}\ \emph {et~al.}(2020)\citenamefont
  {D'Amour}, \citenamefont {Srinivasan}, \citenamefont {Atwood}, \citenamefont
  {Baljekar}, \citenamefont {Sculley},\ and\ \citenamefont
  {Halpern}}]{d2020fairness}%
  \BibitemOpen
  \bibfield  {author} {\bibinfo {author} {\bibfnamefont {A.}~\bibnamefont
  {D'Amour}}, \bibinfo {author} {\bibfnamefont {H.}~\bibnamefont {Srinivasan}},
  \bibinfo {author} {\bibfnamefont {J.}~\bibnamefont {Atwood}}, \bibinfo
  {author} {\bibfnamefont {P.}~\bibnamefont {Baljekar}}, \bibinfo {author}
  {\bibfnamefont {D.}~\bibnamefont {Sculley}}, \ and\ \bibinfo {author}
  {\bibfnamefont {Y.}~\bibnamefont {Halpern}},\ }\bibfield  {title} {\enquote
  {\bibinfo {title} {Fairness is not static: deeper understanding of long term
  fairness via simulation studies},}\ }in\ \href@noop {} {\emph {\bibinfo
  {booktitle} {Proceedings of the 2020 Conference on Fairness, Accountability,
  and Transparency}}}\ (\bibinfo {year} {2020})\ pp.\ \bibinfo {pages}
  {525--534}\BibitemShut {NoStop}%
\bibitem [{\citenamefont {K{\"o}nig}\ and\ \citenamefont
  {Tessone}(2011)}]{konig2011network}%
  \BibitemOpen
  \bibfield  {author} {\bibinfo {author} {\bibfnamefont {M.~D.}\ \bibnamefont
  {K{\"o}nig}}\ and\ \bibinfo {author} {\bibfnamefont {C.~J.}\ \bibnamefont
  {Tessone}},\ }\bibfield  {title} {\enquote {\bibinfo {title} {Network
  evolution based on centrality},}\ }\href@noop {} {\bibfield  {journal}
  {\bibinfo  {journal} {Physical Review E}\ }\textbf {\bibinfo {volume} {84}},\
  \bibinfo {pages} {056108} (\bibinfo {year} {2011})}\BibitemShut {NoStop}%
\bibitem [{\citenamefont {K{\"o}nig}, \citenamefont {Tessone},\ and\
  \citenamefont {Zenou}(2014)}]{konig2014nestedness}%
  \BibitemOpen
  \bibfield  {author} {\bibinfo {author} {\bibfnamefont {M.~D.}\ \bibnamefont
  {K{\"o}nig}}, \bibinfo {author} {\bibfnamefont {C.~J.}\ \bibnamefont
  {Tessone}}, \ and\ \bibinfo {author} {\bibfnamefont {Y.}~\bibnamefont
  {Zenou}},\ }\bibfield  {title} {\enquote {\bibinfo {title} {Nestedness in
  networks: A theoretical model and some applications},}\ }\href@noop {}
  {\bibfield  {journal} {\bibinfo  {journal} {Theoretical Economics}\ }\textbf
  {\bibinfo {volume} {9}},\ \bibinfo {pages} {695--752} (\bibinfo {year}
  {2014})}\BibitemShut {NoStop}%
\bibitem [{\citenamefont {Mariani}\ \emph {et~al.}(2019)\citenamefont
  {Mariani}, \citenamefont {Ren}, \citenamefont {Bascompte},\ and\
  \citenamefont {Tessone}}]{mariani2019nestedness}%
  \BibitemOpen
  \bibfield  {author} {\bibinfo {author} {\bibfnamefont {M.~S.}\ \bibnamefont
  {Mariani}}, \bibinfo {author} {\bibfnamefont {Z.-M.}\ \bibnamefont {Ren}},
  \bibinfo {author} {\bibfnamefont {J.}~\bibnamefont {Bascompte}}, \ and\
  \bibinfo {author} {\bibfnamefont {C.~J.}\ \bibnamefont {Tessone}},\
  }\bibfield  {title} {\enquote {\bibinfo {title} {Nestedness in complex
  networks: Observation, emergence, and implications},}\ }\href@noop {}
  {\bibfield  {journal} {\bibinfo  {journal} {Physics Reports}\ } (\bibinfo
  {year} {2019})}\BibitemShut {NoStop}%
\bibitem [{\citenamefont {Smith}\ and\ \citenamefont
  {Rand}(2018)}]{smith2018simulating}%
  \BibitemOpen
  \bibfield  {author} {\bibinfo {author} {\bibfnamefont {E.~B.}\ \bibnamefont
  {Smith}}\ and\ \bibinfo {author} {\bibfnamefont {W.}~\bibnamefont {Rand}},\
  }\bibfield  {title} {\enquote {\bibinfo {title} {Simulating macro-level
  effects from micro-level observations},}\ }\href@noop {} {\bibfield
  {journal} {\bibinfo  {journal} {Management Science}\ }\textbf {\bibinfo
  {volume} {64}},\ \bibinfo {pages} {5405--5421} (\bibinfo {year}
  {2018})}\BibitemShut {NoStop}%
\bibitem [{\citenamefont {Schweitzer}(2018)}]{schweitzer2018sociophysics}%
  \BibitemOpen
  \bibfield  {author} {\bibinfo {author} {\bibfnamefont {F.}~\bibnamefont
  {Schweitzer}},\ }\bibfield  {title} {\enquote {\bibinfo {title}
  {Sociophysics},}\ }\href@noop {} {\bibfield  {journal} {\bibinfo  {journal}
  {Physics Today}\ }\textbf {\bibinfo {volume} {71}},\ \bibinfo {pages}
  {40--46} (\bibinfo {year} {2018})}\BibitemShut {NoStop}%
\end{thebibliography}%

\end{document}